\documentclass[twocolumn]{aastex63}

\usepackage{amsmath}
\usepackage{pgfplots}
\pgfplotsset{compat=1.15}
\usepackage{tikzscale}
\usepackage{censor}
\usepackage{xfrac}
\usepackage{anyfontsize}
\usepackage{silence}
\usepackage{placeins}
\usepackage{booktabs}
\usepackage{microtype}
\usepackage[caption=false]{subfig}
\usepackage{makecell}

\usepackage{siunitx}
\usepackage[noabbrev,capitalize]{cleveref}
\usepackage{glossaries-extra}
\glsdisablehyper
\makeatletter
\usepackage{etoolbox}
\patchcmd\H@refstepcounter{\protected@edef}{\protected@xdef}{}{}
\makeatother
\newcommand{\eg}{\textit{e.g.}}
\newcommand{\ie}{\textit{i.e.}}
\newcommand{\approximately}{\raisebox{0.5ex}{\texttildelow}}
\definecolor{light-gray}{rgb}{0.8,0.8,0.8}
\def\censorcolor{light-gray}
\let\svcensorrule\censorrule
\renewcommand\censorrule[1]{%
    \textcolor{\censorcolor}{\svcensorrule{#1}}}

\newcommand{\transitDepth}{d}
\newcommand{\planetRadius}{r_{p}}
\newcommand{\targetRadius}{r_{t}}
\newcommand{\targetBackgroundContaminationRatio}{c}

\newcommand{\usraAffiliationString}{Universities Space Research Association, Columbia, MD 21046, USA}
\newcommand{\gsfcAffiliationString}{NASA Goddard Space Flight Center, Greenbelt, MD 20771, USA}
\newcommand{\cuaAffiliationString}{Department of Physics, The Catholic University of America, Washington, DC 20064, USA}

\newcommand{\numberOfHumanVettedCandidates}{181}
\newcommand{\easyLightCurveTic}{258920431}
\newcommand{\hardLightCurveTic}{394346745}
\newcommand{\numberOfConditionalTargets}{8}
\newcommand{\numberOfConditionalCandidates}{4}
\newcommand{\conditionalPairsList}{(TIC 372596795, TIC 372596796), (TIC 120232318, TIC 120232321), (TIC 360816293, TIC 360816296), and (TIC 452810326, TIC 452810327)}
 
\usetikzlibrary{math}
\tikzmath
{
    function symlog(\x,\a){
        \yLarge = ((\x>\a) - (\x<-\a)) * (ln(max(abs(\x/\a),1)) + 1);
        \ySmall = (\x >= -\a) * (\x <= \a) * \x / \a ;
        return \yLarge + \ySmall ;
    };
    function symexp(\y,\a){
        \xLarge = ((\y>1) - (\y<-1)) * \a * exp(abs(\y) - 1) ;
        \xSmall = (\y>=-1) * (\y<=1) * \a * \y ;
        return \xLarge + \xSmall ;
    };
}

 \setabbreviationstyle[acronym]{long-short-user}

\newacronym{Ramjet}{RAMjET}{RApid MachinE-learned Triage}
\newacronym[user1={\citealp{ricker2014transiting}}]{TESS}{\textit{TESS}}{the \textit{Transiting Exoplanet Survey Satellite}}
\newacronym{SNR}{SNR}{signal-to-noise ratio}
\newacronym{LC}{LC}{light curve}
\newacronym{FFI}{FFI}{full-frame image}
\newacronym{MOA}{MOA}{the Microlensing Observations in Astrophysics collaboration}
\newacronym{NN}{NN}{neural network}
\newacronym[user1={\eg, \citealp{lecun2015deep}}]{DNN}{DNN}{deep neural network}
\newacronym[user1={\citealp{krizhevsky2012imagenet}}]{CNN}{CNN}{convolutional neural network}
\newacronym[user1={\citealp{glorot2011deep}}]{ReLU}{ReLU}{rectified linear unit}
\newacronym{ML}{ML}{machine learning}
\newacronym[user1={\citealp{kruse2019detection}}]{QATS}{\texttt{QATS}}{the Quasi-periodic Automated Transit Search pipeline}
\newacronym[user1={\citealp{kostov2019discovery}}]{DAVE}{\texttt{DAVE}}{the Discovery and Vetting of Exoplanets pipeline}
\newacronym[user1={\citealp{brown2018gaia}}]{Gaia}{Gaia}{the Gaia Mission}
\newacronym[user1={\citealp{stassun2018tess}}]{TIC}{TIC}{\textit{TESS} input catalog}
\newacronym[user1={\citealp{exofoptess}}]{ExoFOP}{ExoFOP-\textit{TESS}}{the Exoplanet Follow-up Observing Program for \textit{TESS}}
 
\begin{document}
    \title{Identifying Planetary Transit Candidates in TESS Full-Frame Image Light Curves\\via Convolutional Neural Networks}

% Author information.
\author[0000-0001-8472-2219]{Greg Olmschenk}
\affiliation{\gsfcAffiliationString}
\affiliation{\usraAffiliationString}
\author[0000-0003-2267-1246]{Stela Ishitani Silva}
\affiliation{\gsfcAffiliationString}
\affiliation{\cuaAffiliationString}
\author[0000-0002-3042-4539]{Gioia Rau}
\affiliation{\gsfcAffiliationString}
\affiliation{\cuaAffiliationString}
\author{Richard K. Barry}
\affiliation{\gsfcAffiliationString}
\author[0000-0002-0493-1342]{Ethan Kruse}
\affiliation{\gsfcAffiliationString}
\affiliation{\usraAffiliationString}
\author{Luca Cacciapuoti}
\affiliation{Department of Physics ``Ettore Pancini", Universita di Napoli Federico II, Compl. Univ. Monte S. Angelo, 80126 Napoli, Italy}
\author[0000-0001-9786-1031]{Veselin Kostov}
\affiliation{\gsfcAffiliationString}
\author[0000-0003-0501-2636]{Brian P. Powell}
\affiliation{\gsfcAffiliationString}
\author{Edward Wyrwas}
\affiliation{\gsfcAffiliationString}
\affiliation{Science Systems and Applications, Inc., Lanham, MD 20706, USA}
\author[0000-0002-2942-8399]{Jeremy D. Schnittman}
\affiliation{\gsfcAffiliationString}
\author[0000-0001-7139-2724]{Thomas Barclay}
\affiliation{\gsfcAffiliationString}
\affiliation{University of Maryland, Baltimore County, 1000 Hilltop Circle, Baltimore, MD 21250, USA}

\received{2021 January 26}
\revised{2021 March 23}
\accepted{2021 April 01}
\published{2021 May 21}
\submitjournal{The Astronomical Journal}
     \begin{abstract}
The \textit{Transiting Exoplanet Survey Satellite} (\textit{TESS}) mission measured light from stars in \approximately75\% of the sky throughout its two year primary mission, resulting in millions of \textit{TESS} 30-minute cadence light curves to analyze in the search for transiting exoplanets.
To search this vast data trove for transit signals, we aim to provide an approach that both is computationally efficient and produces highly performant predictions. This approach minimizes the required human search effort.
We present a convolutional neural network, which we train to identify planetary transit signals and dismiss false positives. To make a prediction for a given light curve, our network requires no prior transit parameters identified using other methods. Our network performs inference on a \textit{TESS} 30-minute cadence light curve in \approximately5ms on a single GPU, enabling large scale archival searches.
We present \numberOfHumanVettedCandidates{} new planet candidates identified by our network, which pass subsequent human vetting designed to rule out false positives. Our neural network model is additionally provided as open-source code for public use and extension.
\end{abstract}
     \section{Introduction}
Astronomical photometric data sets are growing at an accelerated pace. Due to their sheer scale, these collections contain data that no human eye has ever nor may ever see. The importance of automated systems, which can filter out data irrelevant to a particular research goal and flag the most promising phenomena, is essential in the era of big data.

The primary goal of \gls{TESS} mission is detecting planets orbiting stars via transit signals in flux measurements. Launched in April 2018, \gls{TESS} is performing a near all-sky photometric survey intended to identify planets with bright enough host stars to enable mass estimation from ground-based radial velocity measurements~\citep{ricker2014transiting}. 

\gls{TESS} is positioned in a high-Earth, 13.7-day, elliptical orbit. In July 2020, \gls{TESS} completed its 2-year primary mission and entered into its extended mission. During the 2-year primary mission, \gls{TESS} recorded measurements of over 200,000 stars at 2-minute cadence. Of more importance for this work, \gls{TESS} recorded flux measurements of its entire field of view (\ang{24}\texttimes\ang{96}) at a 30-minute cadence. This \gls{FFI} data covers \approximately75\% of the sky and provides flux measurements of millions of stars~\citep{ricker2014transiting}.

Along with an abundance of potential transit candidates, \gls{TESS}'s \gls{FFI} dataset presents a challenge: searching the vast dataset in both an accurate and time-efficient way is not trivial. \Gls{ML} generally, and \glspl{NN} specifically, present a solution to this data filtering matter.

In recent years, \glspl{DNN} have come to dominate the field of \gls{ML}. A primary reason for this is that \glspl{NN} have the potential to approximate any transformation function~\citep{cybenko1989approximation,leshno1993multilayer,zhou2020universality}. In this case, we aim to produce a transformation that converts observed data to the true physical classification. Any algorithm (handcrafted or machine-learned) can only approximate such a transformation. When a \gls{NN} is trained for such a task it attempts to learn the optimal data transformations for that specific task~\citep{rumelhart1985learning}. This has the potential to produce significantly more accurate classifications than handcrafted approaches, which often discard valuable information. \Glspl{LC} pre-processed for exoplanet hunting are often detrended using a general Gaussian process to remove stellar variability and other noise sources~\citep{luger2016everest}. Then, a box-fitting least squares algorithm~\citep[\eg,][]{kovacs2002box} is often used on the detrended \gls{LC} to search for transit signals. These processes have to tread a fine line between removing sources of noise and keeping useful signals. Excessive detrending can remove transit signals, while insufficient detrending leads to an abundance of false positive transit detections. Furthermore, the removal of non-transit signals can often be detrimental to correctly identifying a planetary transit signal. For example, one of the most common sources of false positive planet transit signals are eclipsing binaries~\citep{armstrong2016transit}, which have a transit signal similar to a planetary transit candidate. However, eclipsing binary \glspl{LC} also often exhibit an ellipsoidal signal, which differentiates them from a planetary transit~\citep{kostov2019discovery}. Gaussian process detrending often attempts to remove any form of periodic signal that is a non-planetary transit signal, including the eclipsing binary ellipsoidal signal~\citep{foreman2017fast}. This may leave only the eclipsing binary transit signal to be falsely detected as a planetary transit signal.

On the contrary, \glspl{NN} do not explicitly remove sources of noise, but instead learn to use sources of noise to determine the likelihood that a given \gls{LC} contains the desired transit signal, allowing the \glspl{NN} to be effective for noisy data~\citep{dong2014learning,hinton2012deep,xu2014deep}. In the example above, rather than learning to remove the eclipsing binary ellipsoidal signal, the \gls{NN} has the potential to learn that a transit signal with such an ellipsoidal signal likely originates from an eclipsing binary rather than a planet orbiting its host star.

As generalized function approximators~\citep{cybenko1989approximation,leshno1993multilayer,zhou2020universality}, a sufficiently large \gls{NN} can learn any handcrafted transformation, such as the above detrending and box-fitting algorithm. Moreover, if there is a modification to the handcrafted transformation producing better results according to the training process, the \gls{NN} will instead learn that improved transformation. This property of \glspl{NN} gives them the potential to out-perform their handcrafted counterpart in nearly every situation.

This article is organized as follows: in \cref{sec:observational-data} we present the photometric data used in our work. \cref{sec:neural-network-pipeline} presents our \gls{NN} pipeline, including the \gls{NN} architecture (\cref{subsec:network-architecture}) and the pre- and post-processing of the data (\cref{subsec:pre-processing,subsec:post-processing}). We also present the motivation of the network and processing choices in these sections. \cref{sec:results} shows the new vetted planet candidates identified by our network. \cref{sec:results} also provides analysis and dicussion of the new planet candidate population. We conclude the work in \cref{sec:conclusion}.     \section{Observational Data}\label{sec:observational-data}
\begin{figure*}
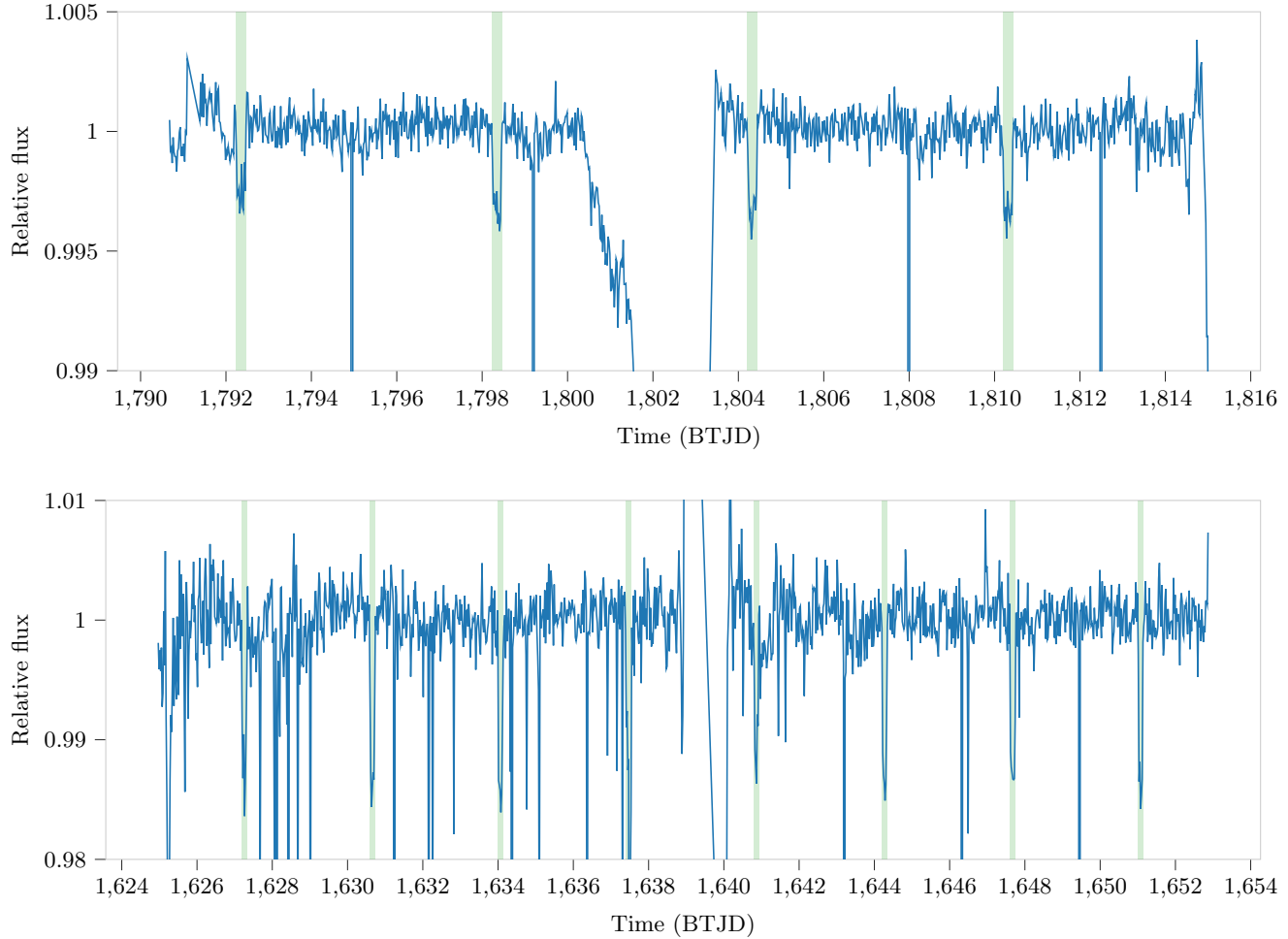

    \centering
    \subfloat{
    \includegraphics[width=0.98\textwidth, height=0.35\textwidth]{easy_example_light_curve.tikz}
    }
    \\
    \subfloat{
    \includegraphics[width=0.98\textwidth, height=0.35\textwidth]{hard_example_light_curve.tikz}
    }
    \caption{Examples of \gls{TESS}  30-minute cadence single sector \gls{LC} data from \gls{TESS} observations for TIC \easyLightCurveTic{} (top panel) and TIC \hardLightCurveTic{} (bottom panel). Time is in $\text{BTJD} = \text{BJD}-2457000.0$. The dips in flux (highlighted by green vertical bars) are caused by the transits of planet candidates identified by our network.}
    \label{fig:light_curve_example}
\end{figure*}

For the present work we have used \Gls{LC} data, \ie, measures of flux over time. \cref{fig:light_curve_example} illustrates an example of a \gls{TESS} \gls{LC}, showing time vs.~flux of two \gls{TESS} targets: TIC \easyLightCurveTic{} and TIC \hardLightCurveTic{}. The time is given in \gls{TESS} Barycentric Julian Day (BTJD) time~\citep{tenenbaum2018tess}. With the Julian Day in the Barycentric Dynamical Time standard (BJD), $\text{BTJD} = \text{BJD}-2457000.0$. BJD is usually the most accurate time standard to use as it accounts for many different timing corrections, including leap seconds~\citep[\eg,][]{eastman2010achieving}. The flux given is median normalized flux for the \gls{LC}. The \gls{LC} was produced by the \texttt{eleanor} pipeline~\citep{feinstein2019eleanor} from raw flux measurements provided by \gls{TESS}.

The examples shown in \cref{fig:light_curve_example} are planetary candidates identified by our \gls{NN}. We selected these \glspl{LC} to show cases of relatively simple noise and relatively challenging noise in our dataset. These \glspl{LC} demonstrate some typical sources of noise and data incompleteness in the \gls{TESS} FFI data. Indeed, the repeating sharp downward spikes and the more gradual spikes near the start and end of an observing session are caused by spacecraft systematics and/or detrending; there is a gap in the middle of the data caused by the spacecraft pausing its observing to downlink data to Earth~\citep{tenenbaum2018tess}.

Ideally, a \gls{LC} would contain only the flux from a single \gls{TESS} target (typically a star system). However, in reality each \gls{TESS} pixel covers \approximately21 arcseconds of the sky, and \gls{TESS}'s point spread function results in blending between pixel measurements. For these reasons, a LC will contain flux from multiple targets. This often makes it challenging to determine which source the signal (or noise) is originating from.

\subsection{\textit{TESS} data}
The \gls{TESS} data sets include 2-minute cadence \gls{LC} data and 30-minute cadence \gls{LC} data, both of which are relevant in this work. In the following sections, we describe each of these data sets and their use in our study.

\subsubsection{\textit{TESS} 2-minute cadence \glsfmtlong{LC} data}
\label{subsubsec:tess-2-minute-cadence-lc-data}
\Gls{TESS} takes measurements of a large portion of the sky at regular intervals. During the primary mission, this interval was every 2 minutes. However, due to limitations of the spacecraft's storage and downlinking capabilities, only a small portion of this 2-minute cadence was retained~\citep{ricker2014transiting}. For the present work, the set of known planets and planet candidates we employ for training our NN comes primarily from searches into the 2-minute cadence data set. For each 2-minute cadence measurement, the data come from small patches of pixels around targets likely to be of interest in that portion of the sky; a description of the selection criteria for \gls{TESS} targets can be found in \citet{stassun2018tess}. The pixels within these patches that are suspected to contain a high \gls{SNR} are then summed to form a single flux measurement of a \gls{LC}. Such summed fluxes are combined for each 2-minute cadence measurement, forming the content of the 2-minute cadence \gls{LC}s we refer to in this work.

\gls{TESS} collected \approximately600,000 2-minute cadence \glspl{LC} from \approximately200,000 targets during its primary mission~\citep{tenenbaum2018tess}. Of these targets, \approximately326 exhibit the transit signal of a confirmed planet~\citep{exofoptess}.

\subsubsection{\textit{TESS} 30-minute cadence \glsfmtlong{LC} data}\label{subsec:30-minute-cadence-data}
\gls{TESS} discards most of the pixels from the 2-minute cadence measurements (see \cref{subsubsec:tess-2-minute-cadence-lc-data}). However, at a 30-minute cadence all pixels' values are retained and downlinked to Earth. These \glspl{FFI} cover a much larger number of targets at a lower time resolution~\citep{tenenbaum2018tess}. We used \approximately67 million 30-minute cadence \glspl{LC} (with \gls{TESS} magnitudes \textless15) in this work, and this is the primary data set investigated by our \gls{NN}.
     \section{Neural network pipeline}\label{sec:neural-network-pipeline}
\begin{figure*}
    \centering
    \includegraphics[width=0.9\textwidth]{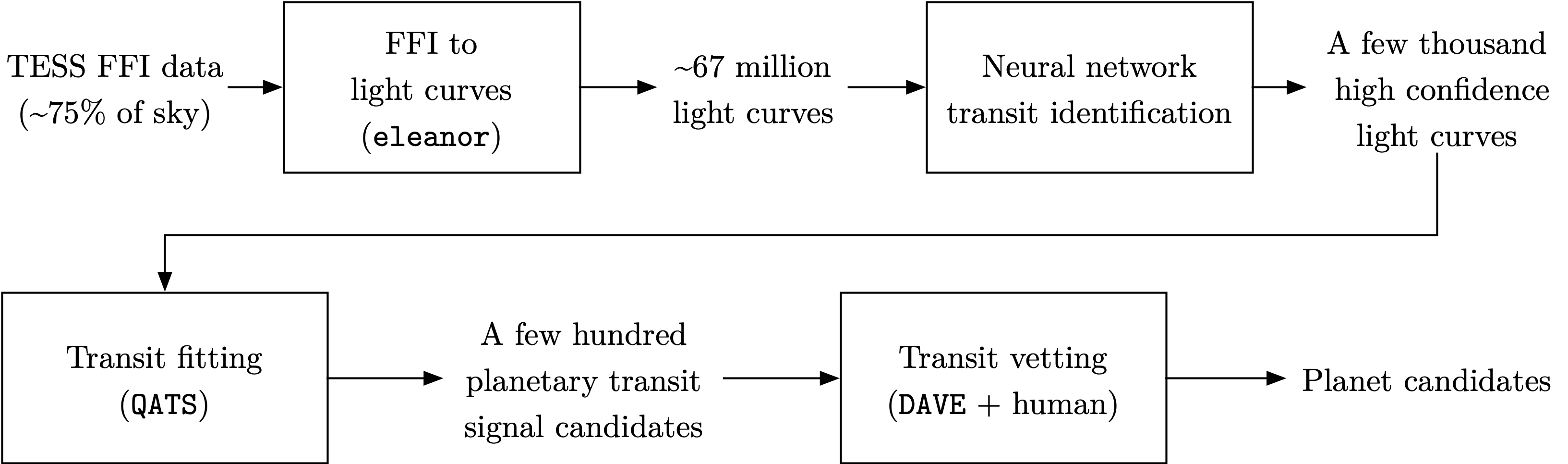}  
    \caption{A conceptual overview of our data pipeline. The new contributions in this work are primarily those related to the development and use of of the neural network. We briefly describe other steps of the full pipeline in this work.}
    \label{fig:conceptual_pipeline}
\end{figure*}
A conceptual overview of our pipeline is shown in \cref{fig:conceptual_pipeline}. The new contributions of this work focus on development and use of the \gls{NN} portion of this pipeline (including data pre-processing). Brief descriptions of the other pipeline components are given below. For details on the other components of the data pipeline see respectively: Kruse et~al. (2021, in preparation) for the production of \glspl{LC} from \gls{TESS} FFIs (via \texttt{eleanor}~\citealp{feinstein2019eleanor}); \citet{kruse2019detection} for \glsentrylong{QATS} (\glsentryshort{QATS}); and \citet{kostov2019discovery} for \glsentrylong{DAVE} (\glsentryshort{DAVE}).

\subsection{Neural network primer}
NNs are generalized transformation function learning machines. In our case, the LC is represented as an array of flux values, and the transformation function the network is learning is a transformation from the LC to a prediction of the likelihood that the LC contains a planetary transit signal. The transformation learned by a NN has two notable properties. First, a sufficiently large NN can approximate any transformation, including any handcrafted transformation~\citep{cybenko1989approximation,leshno1993multilayer,zhou2020universality}. Second, the transformation function is automatically learned based on training examples. To train a network, we ``show" the network examples of LCs known to contain a planetary transit signal and those known to \textit{not} contain such a signal. We then ``ask" the network which examples contain planetary transits and which do not. At the start of the training, the network will make predictions similar to random guesses. We use the network's confidence of each prediction to update the parameters of the network's transformation function~\citep{rumelhart1986learning}. Each parameter in the network is updated with a small change to produce a slightly better prediction for the LC(s) we are currently showing to it. The direction and magnitude of these parameter updates are determined through backpropagation~\citep{rumelhart1986learning}, which determines the derivative of the NN's parameters with regard to its prediction's correctness. By repeating these prediction and update steps many times, the network approaches a transformation function that can distinguish between LCs with and without planetary transit signals.

\subsection{Design choice overview}
\Gls{ExoFOP} has confirmed only a few hundred planetary targets within \gls{TESS} \gls{LC} data. Typically, this number of examples is too small to train a NN of the scale we are using without overfitting to the training data. However, we have layered several techniques to prevent overfitting and promote generalization. These include data augmentation (\cref{subsec:pre-processing}), data generation through injection (\cref{subsec:ground-truth-training-dataset}), and various network mechanisms (\cref{subsec:network-architecture}).

NN overfitting can be intuitively understood by considering the example of a large network with a small amount of training data. In such a case, the NN has the potential to memorize the training examples. For example, such a NN may learn to look exclusively at the first flux value of each LC. So long as this first flux value is unique for each LC, the NN can simply learn which unique values correspond to planetary transit examples and which correspond to non-planetary transit examples. This produces a NN model that will perfectly distinguish between examples in the training data but will do no better than random guessing when applied to new data. This, however, is only one extreme case of overfitting. The NN may overfit based on several other LC features or statistics. Preventing such overfitting is one of the primary reasons we introduce the various augmentations and mechanisms below~\citep[\eg,][]{srivastava2014dropout,ioffe2015batch}.

When designing and training the network, we use 80\% of the \gls{ExoFOP} confirmed planets. We use another 10\% as validation data. These are data set aside that the network is not trained with. Instead, these data are used to evaluate the predictive performance of the trained network. This means inspecting how correctly the network makes predictions on data the network was not trained with, but where we know the correct answer. The remaining 10\% is set aside as test data to evaluate the trained network after all design decisions are finalized. Several of the specific network and training setup design decisions were guided by preliminary performance results on the validation data. However, this validation evaluation and the evaluation on the test data are beyond the scope of this work. A detailed evaluation of the various network mechanisms and training techniques used in this work (and several excluded from this work) will be provided in Olmschenk et~al. (2021, in preparation).

Throughout the following sections describing the network and related processing, we provide subsections explaining the rationale behind the design choices.

\subsection{Network architecture}\label{subsec:network-architecture}
\begin{figure*}
    \centering
    \begin{minipage}{0.4\textwidth}
        \centering
        \subfloat[The full network.\label{fig:ffi_hades_network}]{
            \centering
            \includegraphics[width=0.85\textwidth]{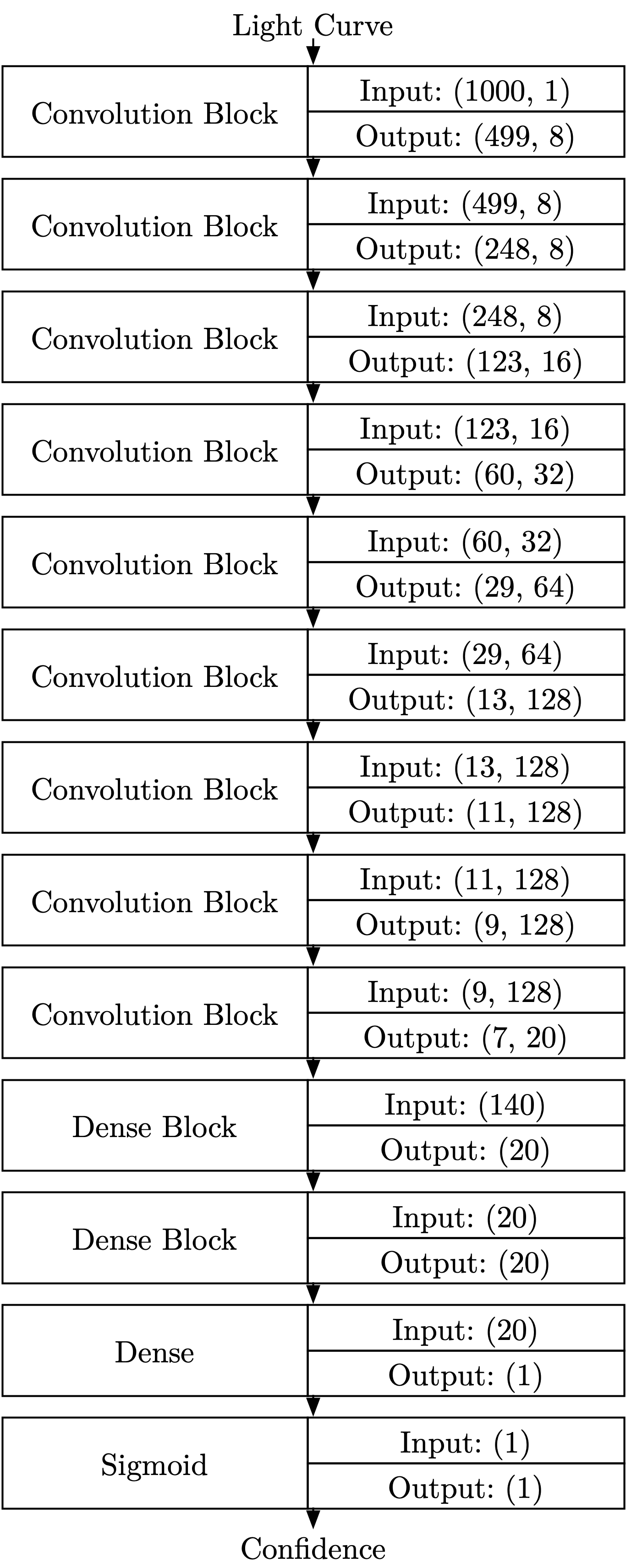}
        }
    \end{minipage}
    \begin{minipage}{0.25\textwidth}
        \vfill
        \centering
        \subfloat[The outline of a convolution block structure.\label{fig:ffi_hades_convolution_block}]{
            \centering
            \includegraphics[width=0.65\textwidth]{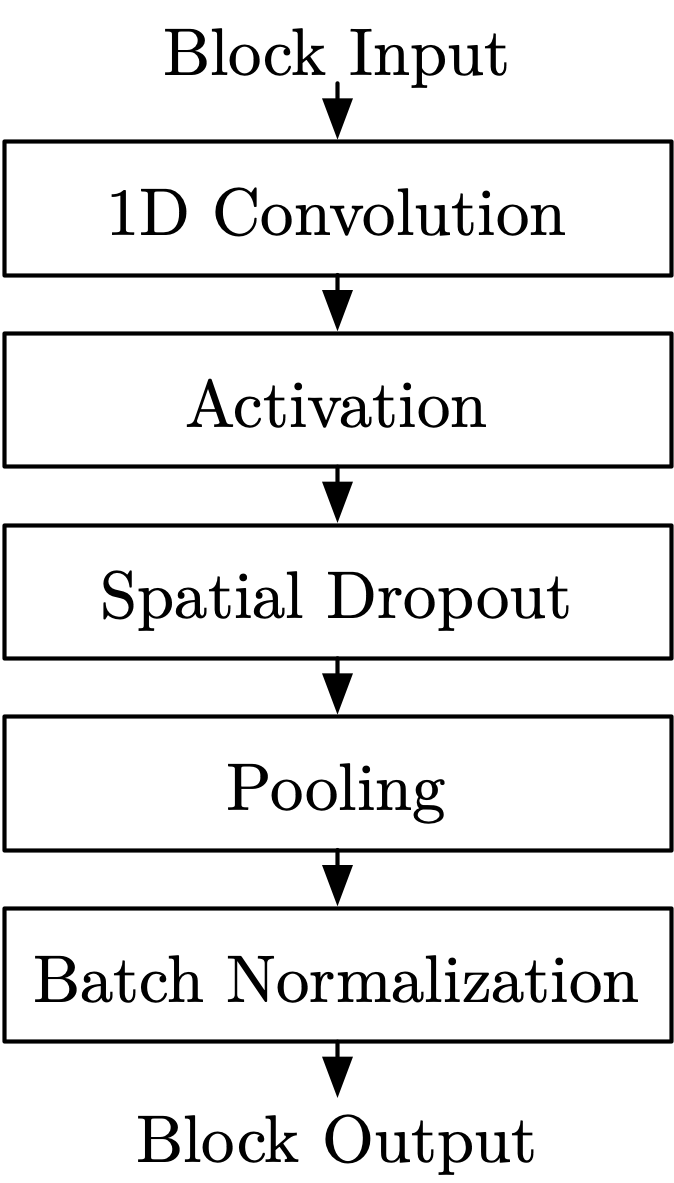}
        }
        \\
        \vspace{1cm}
        \subfloat[The outline of a dense block structure.\label{fig:ffi_hades_dense_block}]{
            \centering
            \includegraphics[width=0.65\textwidth]{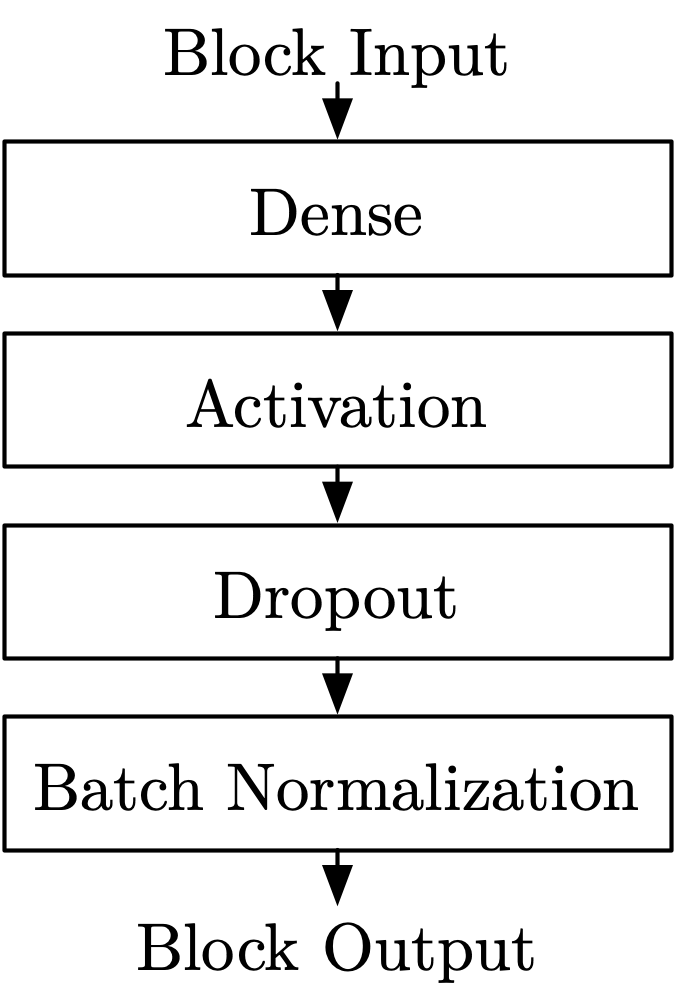}
        }
        \vfill
    \end{minipage}
    \caption{The general architecture of the convolutional neural network used in this work. See text
    for details. All convolution/dense layers within a block use a number of filters/units equivalent to the size of
    the last dimension of their output tensor. All convolutions use a kernel size of 3. To preserve the clarity of
    the diagram, three inconsistencies in network blocks are not shown. First, the first convolution block and the last
    dense block do not apply dropout or batch normalization. Second, the final convolution block applies a standard
    dropout instead of spatial dropout, as the following layer is a dense layer. Finally, only the first 6
    convolution block use pooling (with pooling size 2). The remaining convolution blocks do not pool.}
    \label{fig:network_diagram}
\end{figure*}

In this work we used the 1D \gls{CNN}, which is shown in \cref{fig:network_diagram}.
 All layers use leaky \gls{ReLU} activations, except the final prediction layer, which uses a sigmoid activation~\citep{wilson1972excitatory}. Excluding the first and the last two, all layers apply dropout~\citep{srivastava2014dropout} and batch normalization~\citep{ioffe2015batch}. The dropout rate is $0.1$ and is enabled during training and disabled during inference. We use spatial dropout~\citep{tompson2015efficient} for any convolutional layers. This is a version of dropout better suited for convolutional layers that drops entire features as opposed to individual neurons~\citep{tompson2015efficient}. Batch normalization moving averages are updated only during training~\citep{srivastava2014dropout}. We used max pooling following~\cite{krizhevsky2012imagenet}. The ordering of the components in all layers is convolution/dense transformation, activation, dropout, pooling, and batch normalization. We used an Adam optimizer~\citep{kingma2014adam} on a binary cross-entropy loss to train the network. The output of the network for a given input \gls{LC} is the network's confidence that the \gls{LC} contains a planetary transit signal. We note that these are uncalibrated confidences, and therefore the distribution of the network's confidences may not directly correspond to the true underlying physical distribution.

The \gls{NN} pipeline code is available at \url{https://github.com/golmschenk/ramjet} (see \citealt{greg_olmschenk_2021_4460662} for the code version used in this work). This pipeline is also installable as a PyPI package (\url{https://pypi.org/project/astroramjet/}). Documentation for the \gls{NN} pipeline can be found at \url{https://astroramjet.readthedocs.io/en/latest/}.

\subsubsection{Network architecture rationale}\label{subsubsec:network-architecture-rationale}
We chose a 1D \gls{CNN} over other alternatives, such as dense networks, for the following reasons. First, we expect the \gls{NN} to find individual transit events as a primary feature. In early layers, the \gls{NN} should ignore the position of the transit in the \gls{LC}, and only determine their presence based on the local \gls{LC} shape. As such, the early layers of our network are convolutional layers, which treat each segment of the \gls{LC} identically~\citep{krizhevsky2012imagenet}, \eg, they search each portion of the LC for a transit occurring in that location. Only after local level features, \eg, individual transits are discovered, do we expect the network to combine these features into global level features---in this case repeating periodic transits. For this reason, in the NN the convolutional layers are followed by global dense layers. Another advantage of convolutional layers is to prevent overfitting; this happens because weights for convolutional layers are shared on every part of the input, preventing the network from applying specific weights to specific positions in the \gls{LC}~\citep{krizhevsky2012imagenet}. Our \gls{NN} requires no prior transit parameter information; the only input to the network are the flux values of the \glspl{LC}. The \gls{NN} performs inference on a \gls{LC} in \approximately5ms on a single GPU, allowing the entire data set of \approximately67M \glspl{LC} to be inferred on in a few days.

A common source of overfitting comes from the NN identifying a specific training example, or set of training examples, based on a limited number of features that uniquely identify them; for example, the combination of the first and second flux values, which are floating point values, may be unique for every \gls{LC}. This allows the network to distinguish training examples, but these unique features do not generalize to data the network did not train with. We added dropout to prevent such overfitting~\citep{srivastava2014dropout}. For each layer where it is applied, dropout randomly sets feature activations to zero during training. This prevents the network from relying on a small number of features to determine the network confidence that specific training \glspl{LC} contain transits. As the network can no longer rely on specific features to exist that uniquely identify training examples, the network is encouraged to produce features that detect transits, \ie, the shared feature contained in the positive training data. During inference of non-training data, dropout is disabled to allow the network to use all features.

We apply batch normalization both to improve training dynamics and prevent overfitting. On each layer where it is applied, batch normalization normalizes the activations of the layer for the given batch of training data. This aids in deep network training, as it provides well-distributed training gradients, thereby avoiding the vanishing/exploding gradient problem~\citep{hochreiter1998vanishing}. Applying batch normalization also prevents overfitting, as each normalization depends on the batch of data and on the current network weights. As each normalization affects the input of the network layer, even small batch or weight changes have a cascading effect throughout the network. This makes it difficult for the network to overfit to specific \glspl{LC}, and encourages the network to converge toward a generalized solution~\citep{ioffe2015batch}.

\subsection{\Glsfmtlong{FFI} \glsfmtlong{LC} production}
For details of the \gls{FFI} \gls{LC} production, see Kruse et~al. (2021, in preparation). Briefly, Kruse et~al. (2021, in preparation) used the 129,000-core \textit{Discover} supercomputer at the NASA Center for Climate Simulation, to build \gls{FFI} \glspl{LC} for all stars observed by \gls{TESS} down to 15th magnitude. All original and calibrated FFIs were produced by the \gls{TESS} Science Processing Operations Center~\citep{jenkins2016tess}. Target lists were created through a parallelized implementation of \texttt{tess-point}~\citep{burke2020tess} on the \gls{TIC} provided by the \citet{masttess}. The \glspl{LC} for each sector were constructed in 1--4 days of wall clock time (for a total of over 100 CPU-years), depending on the density of targets in the sector, through a parallelized implementation of the \texttt{eleanor} Python module~\citep{feinstein2019eleanor}. \approximately67 million \glspl{LC} were produced at the time this work was performed. These single sector \glspl{LC} are the input to the pre-processing and, subsequently, our \gls{NN}.

\subsection{Pre-processing}\label{subsec:pre-processing}
We used several forms of data augmentation to prevent network overfitting and to encourage generalization of learned features. In some cases, we did not apply the data augmentation during inference to allow for reproducibility and to allow for the best available input information during the inference phase.

During training, each time a \gls{LC} is prepared for input to the network, the pipeline removes a random subset of the data points. The ratio of data points removed is randomly selected from $\mathcal{U}(0, 0.01)$. The implementation of this removal shifts the remaining values in the array such that there are no gaps. During inference, no data points are removed in this way.

Next, the \gls{LC} is randomly rolled, \ie, a random position is chosen in the \gls{LC} and the data are split at that location. The order of these two pieces is reversed. This rolling is not applied during inference.

Afterward, the pipeline repeats or truncates the \gls{LC} to have a uniform length of 1,000 data points. This is approximately the median length of a single sector \glspl{LC} generated from \gls{TESS} FFI data. \Glspl{LC} shorter than 1,000 data points are repeated, with the first values of the LC being appended to the end of the LC, until they are 1,000 data points. \Glspl{LC} longer than 1,000 data points are truncated. This transformation is applied during both training and inference.

Finally, the flux values of \glspl{LC} are normalized before being input into the network, using a normalization in the following way. A percentile normalization is applied such that the 10$^{\text{th}}$-percentile flux is normalized to $-1$ and the 90$^{\text{th}}$-percentile flux is normalized to $1$. This flux normalization is applied during both training and inference.

\subsubsection{Pre-processing rationale}\label{subsubsec:pre-processing-rationale}
As described above, we performed several data augmentation steps. We have chosen to do so for the following reasons:

\textbf{First}, randomly removing data points during training helps prevent the network from overfitting. Indeed, a large NN has the potential to memorize exact values or ordering of values within the input data. By removing random data points during training, we encourage the network to not rely on specific data points but to use the overall structure of the LC instead~\citep{zhong2020random}. Removing random data points during inference has no benefit and could potentially remove valuable information, so we only removed data points during training.

\textbf{Second}, the random roll of the LCs helps prevent the network from searching for specific positions of features within the \glspl{LC}. Sector-specific noise can be easily memorized by its position in the LC, and rolling the LC forces the network to generalize feature recognition tasks to the general LC structure as opposed to a single part of it. This process splits the \gls{LC} into two pieces and swaps the order of these pieces; therefore the time between two transits where the \gls{LC} are recombined will not match the original period. As we expect, the network will take into account the period when inferring for any given \gls{LC}, so this process may be a slight detriment to the training. However, we determined through preliminary validation experiments that the generalization benefit outweighs the cost. Similar to the random data point removal, there is no benefit to apply this step during inference.

\textbf{Third}, the uniform length of 1,000 data points per \gls{LC} allows for a significantly more efficient and practical training dataset for our pipeline. This is because the network only needs to be designed for a single length input and can process in parallel large batches of uniform length inputs. In principle, this uniform LC length could cause two potential detriments to training: 1) transit events may be excluded when the LC is truncated, and 2) a pair of transit events may be artificially given the incorrect period when the LC data are repeated. However, both these factors play only a minor role in altering the LC before its input into the network. Therefore, we determined that for both cases the benefits of having this uniform LC length outweighed the costs.

\textbf{Fourth and lastly}, the data augmentation of percentile normalization of the flux values provides several benefits. Due to the previous data removal, rolling, truncating, and repeating, the LC is normalized differently in one training step than in another. This provides another deterrent to network overfitting, as the NN cannot rely on exact LC input values to define the LC label. Inputs well distributed from -1 to 1 provide several benefits to internal network training dynamics~\citep{lecun2012efficient} and allow for simplified weight initialization~\citep{glorot2010understanding}. We chose a percentile normalization over a standard normalization, as this results in most of the data points being well distributed from -1 to 1. Notably, this provides better distributed values than it does a standard normalization where the minimum and maximum are scaled to -1 and 1 respectively. This is because several astronomical events result in outlier fluxes, which would result in the majority of a LC's data points being normalized to very near -1 or very near 1 when using a standard normalization . For example, a flare may result in a few flux values being relatively high. The standard normalization results in the non-flare flux values being normalized to close to -1 for all values. In contrast, the percentile normalization results in most values being normalized from -1 to 1, which provides better training conditions for the network.

\subsection{Ground truth training dataset}\label{subsec:ground-truth-training-dataset}
We used the dispositions of \gls{ExoFOP} for ground truth training labels of \gls{LC} transits. Positive cases included \Glspl{LC} corresponding to confirmed planets, according to the \gls{ExoFOP} catalog.  We included as negative cases any targets not listed by \gls{ExoFOP} or that \gls{ExoFOP} designated as false positives. We excluded from the training process any target designated by \gls{ExoFOP} as a not confirmed candidate. This resulted in \approximately377 30-minute cadence \glspl{LC} of targets with known planet transit signals. We used this collection of targets for training, validation, and testing of our NN.

In addition, we used a catalog of eclipsing binaries~(Kruse et~al. 2021, in preparation) as negative cases, as eclipsing binaries are the most likely targets to result in false positives (see \cref{subsubsec:ground-truth-training-dataset-rationale}).

The \gls{NN} was trained using these initial training data sets, and human researchers analyzed the top candidates output by the network.

During training, we showed the network three base sets of \glspl{LC} at equal rates: 1) \glspl{LC} of known transiting planets, 2) \glspl{LC} of targets from an eclipsing binary catalog, and 3) all available non-transit \gls{ExoFOP} candidate \glspl{LC}. In each case, only \gls{TESS} 30-minute cadence \glspl{LC} were used as described in \cref{subsec:30-minute-cadence-data}.

\begin{figure*}
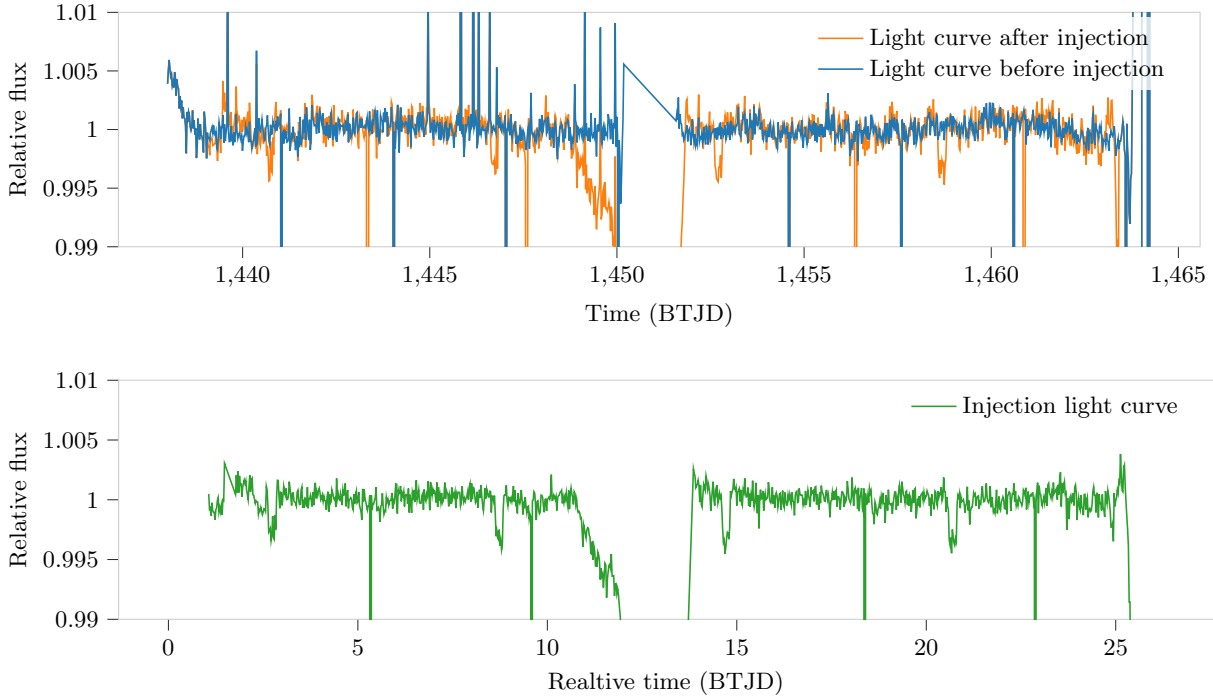

    \centering
    \subfloat{
        \includegraphics[width=0.9\textwidth, height=0.25\textwidth]{injection_explanation_injectee_and_injected.tikz}
    }
    \\
    \subfloat{
        \includegraphics[width=0.9\textwidth, height=0.25\textwidth]{injection_explanation_injectable.tikz}
    }
    \caption{An example of the injection described in the text. The top panel shows a light curve before and after it is injected with another light curve. The bottom panel shows the light curve used in the injection.}
    \label{fig:injection-explanation}
\end{figure*}

We additionally trained the network with LCs artificially injected with signals from another LC. Thus, in addition to the three base \gls{LC} sets above, we trained the network with three injected \gls{LC} sets, one for each of the three base sets. The corresponding injected set for each of the base sets is produced as follows. During training, we randomly sampled a LC from the base set. This LC is median normalized to produce a relative magnification signal. We then randomly sample a LC from the non-transit candidate set, and then multiply each value in this LC by the relative magnification signal generated from the previous LC. An example of this injection is shown in \cref{fig:injection-explanation}. Please note that the non-transit LC set is always used as the source of LCs to have a signal injected into, but is also used for the source of signals to inject in one of the three injected sets.

During this injection process, we interpolated linearly between generated signal times to determine signal magnifications to be injected. LCs injected with signals from the known transit base LC set are labeled with a positive ground truth label. Those injected with signals from the eclipsing binary and non-transit candidate sets are labeled with a negative ground truth label. The resulting LCs produced by this injection process are treated identically to base set LC (\eg, are prepared for input to the network using the same pre-processing steps). We train the network sampling evenly from each of the six LC sets, \ie, using three base sets and three injected sets.

\subsubsection{Ground truth training dataset rationale}\label{subsubsec:ground-truth-training-dataset-rationale}
The known planetary transit signals we use to train our network were identified primarily by non-\gls{NN} search methods working on the 2-minute cadence LC data~\citep{huang2020photometry}. However, we train the network using the corresponding 30-minute cadence LC data. In this way, the \gls{NN} is trained to predict the same labels on lower-quality data that traditional methods obtained using higher-quality data. That is, the network is not simply learning to duplicate the traditional methods; it must learn to identify the same targets with $1/15$th the cadence. This has the potential to have the \gls{NN} to learn patterns that would be missed by traditional algorithms. 

We used two types of negative datasets to train our NN: (1) the dataset of all 30-minute cadence LCs, which are from targets that are not known planetary candidates or confirmed planets; and (2) the dataset of all 30-minute cadence LCs labeled as eclipsing binary candidates according to Kruse et~al. (2021, in preparation). The reasoning for having the two separate sets of negative \glspl{LC} arises from the ratio of occurrences of each type of phenomenon. The vast majority of \gls{TESS} 30-minute cadence \glspl{LC} (\textgreater99\%) are expected to contain neither planetary transit signals nor eclipsing binary signals. Most of the remaining \gls{LC} signals are not due to transiting planets, but instead to eclipsing binaries (\textgreater95\%)~\citep{sullivan2015transiting}.

We excluded from the training process any targets designated by \gls{ExoFOP} as a planet candidate that were not confirmed. The reason behind this choice is that candidates added to \gls{ExoFOP} are frequently subsequently confirmed or designated as false positives, and we want to train our network only with signals of planets~\citep{exofoptess}.

To understand our training setup, we first considered a simpler setup that would sample from all available \glspl{LC} equally, assigning the appropriate label to each \gls{LC}. In this training setup planetary transit \glspl{LC} would be very rare, and the \gls{CNN} would have little incentive to learn to predict them, as predicting negative in every case would provide the correct answer for nearly every \gls{LC}. This would be compounded by frequent mini-batches, which have no positive cases and would then result in training noise.

Next, we considered a training setup using equal cases of negative and positive \glspl{LC}. This forces the \gls{CNN} to learn to distinguish between non-planetary-transit \glspl{LC} and planetary-transit \glspl{LC}. However, in this case there is an issue in training data ratios. The \gls{CNN} can obtain the correct prediction on nearly every \glspl{LC} it is shown simply by labeling any periodic event as positive and any non-periodic event as negative. While this will help to filter out quiescent or otherwise non-periodic \glspl{LC}, the vast majority of \glspl{LC} labeled as positive will have signals not caused by transit events. Instead, they will be most often due to eclipsing binaries, or other periodic signals somewhat resembling short dips in flux.

The solution we employed to handle these labeling imbalances was to train the \gls{CNN} with \glspl{LC} sampled equally from 3 sets of \glspl{LC}: 1) all negatives, 2) eclipsing binaries, and 3) planet transits. In this way, the \gls{CNN} must be able to distinguish non-periodic events from periodic ones, to correctly make predictions about the general negative \glspl{LC}; but it must also be able to distinguish eclipsing binaries from planetary transit events. As eclipsing binaries often look very similar to planetary transits~\citep{kostov2019discovery}, this forces the \gls{CNN} to learn explicitly how a planetary transit appears relative to other types of periodic signals.

%\subsection{Discussion of injected LC data}\label{subsubsec:discussion-of-injected-lc-data}
We produced the artificially injected LCs to provide the network with examples of signals in a variety of real noise. With only a few hundred confirmed training examples, \glspl{NN} may be prone to only learn the specific positive examples provided. By injecting the known signals into other LCs, we force the network to learn to recognize transit signals within any other LC in our training dataset. This encourages the network to learn that the transit signal is the important feature to identify and encourages it to learn how to ignore any other signals.

These artificially created \glspl{LC} retained noise from both the injected and the injectee LCs. This results in statistically more noisy artificial LCs than the average real LC. This could lead the network to wrongly learn to give higher confidence to noisier LCs. To counteract this, we also used injected eclipsing binary and non-transit signals. In order to provide realistic noise cases, we trained with both the injected LCs, and with the original real data LCs; this choice has been made because the real data are expected to have similar amounts of noise to the data the network will perform inference on.

The evaluation of the impact this injection technique has on predictive performance goes beyond the scope of this work and will be presented in Olmschenk et~al. (2021, in preparation).

\subsection{Post-processing}\label{subsec:post-processing}
After the network produces a confidence value for each LC an arbitrary number of the highest confidence candidates are passed through the post-processing portion of the pipeline.

First, the candidate target LCs are passed through \gls{QATS}, which provides the fitting of a transit model for each LC. As with the NN, the input LCs used by \gls{QATS} are produced via the \texttt{eleanor} pipeline~\citep{feinstein2019eleanor}. As \texttt{eleanor} cleans and detrends the LC data, occasionally this process results in a LC with spurious relative flux scales. In particular, in the processed version of the \gls{LC}, the depths of the transits may be artificially reduced. As \gls{QATS} uses these relative flux scales to estimate the depth of the transits, this also affects our radius estimates. Notably, our candidates with the smallest radii likely have underestimated radii. Most of the estimates are expected to be accurate. The transit model determined by \gls{QATS} supplies transit parameters such as transit depth, duration, period, and epoch. These transit parameters are used by the remaining parts of the pipeline.

Based on the transit parameters determined by \gls{QATS}, we filter candidates on the predicted radius $\planetRadius$, calculated as follows:
\begin{equation}
    \planetRadius = \targetRadius \sqrt{\transitDepth(1 + \targetBackgroundContaminationRatio)}
    \text{,}
\end{equation}
where $\targetRadius$ is the target star radius obtained from \gls{Gaia}'s data release 2, 
$\targetBackgroundContaminationRatio$ is the target's background contamination obtained from the \gls{TIC}, and $\transitDepth$ is the transit depth as modeled by \gls{QATS}. The pipeline discards planet candidates with predicted radius greater than $1.8~R_{\text{Jupiter}}$. We chose this threshold to be over the 95\% exoplanet radii expected to be discovered in \gls{TESS} FFI data~\citep{barclay2018revised}, while still being below the radii of largest known exoplanets~\citep{zhou2017hat, crouzet2017discovery}.

The pipeline then passes any candidates that are not removed from the above filtering to \gls{DAVE}, which provides an automated vetting of transit candidates. This includes checking for secondary signals and for in- and out-of-transit difference image photometric centroid shifts.

Finally, a group of exoplanet researchers visually examine \gls{QATS} and \gls{DAVE} analysis results to accept or reject the candidates.

\subsubsection{Post-processing rationale}\label{subsubsec:post-processing-rationale}
The pipeline discarded planet candidates with predicted radius greater than $1.8 \cdot R_{Jupiter}$. This threshold allows more than 95\% of planet radii expected to be found in \gls{TESS} FFI signals~\citep{barclay2018revised}, while being below the radii of the largest known exoplanets~\citep{zhou2017hat, crouzet2017discovery}. Objects with a radius greater than this threshold might be brown dwarfs~\citep{carmichael2020two}.

The final human analysis, done with the results of \gls{QATS} and \gls{DAVE}, consisted of removing any candidates that were likely caused by a non-planetary transit signal. The most common source of such false positives were eclipsing binaries. This occurred most often when a nearby eclipsing binary's signal appeared in the target's LC. Often, this can be seen due to a in/out-of-transit photo difference centroid offset~\citep{kostov2019discovery}. Often targets have another neighboring target a subpixel distance away, where the brightness of the neighboring target was such that an eclipsing binary transit signal from that source would appear as a planetary transit signal, from the primary target. In such ambiguous cases, the candidate was discarded.

\subsection{Active learning}
The positive labels confirmed by the researchers were fed back to the network's training process in order to supplement its list of positive training candidates. Some newly identified eclipsing binary cases were also labeled as negatives and fed back to the training process. We performed this active learning in a subjective fashion, and no formal process was used to guide when or how it should occur. However, we provide here an approximate description of the process. Once the network training had converged, we passed \approximately1,000 candidates with the highest confidence to the post-processing portion of the pipeline. The output of \gls{QATS} typically showed \approximately90\% of these candidates to have unrealistic physical parameters for transiting planets, leaving \approximately100 to be analyzed by \gls{DAVE} and a human researcher. \approximately30\% of these candidates passed the human vetting reviewing the output of \gls{DAVE}, leaving \approximately30 candidates. These candidates were then added back into the training dataset. This process was repeated \approximately6 times.
     \section{Results and analysis}\label{sec:results}

\begin{figure}
    \centering
    \includegraphics[width=\columnwidth]
    {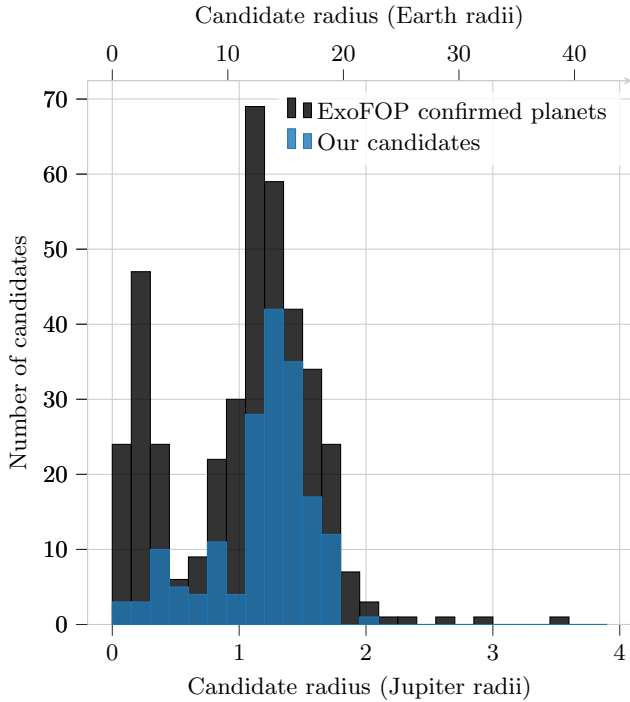}
    \caption{The radii distribution of our planet candidates and the confirmed planets from \gls{ExoFOP}.}
    \label{fig:candidate_radii_distribution_plot}
\end{figure}

The primary output of this work is the human-vetted planet candidates shown in \cref{tab:planet_candidate_table}. These \numberOfHumanVettedCandidates{} candidates have passed the entire automated vetting process and were verified by humans. The radii given in~\cref{tab:planet_candidate_table} are estimated using the method described in \cref{subsec:post-processing}.
The distribution of the candidates' radii is shown in \cref{fig:candidate_radii_distribution_plot}.

In this section, we examine the planet candidates we identified and compare them to the confirmed \gls{ExoFOP} planets. We also compare our findings with the estimations from \citet{barclay2018revised} on the expected exoplanet yield of the \gls{TESS} mission and expected physical properties of the population of exoplanets and their host stars. Follow-up analyses, especially radial velocity measurements, are necessary to confirm our candidates as planets; however, we compare various properties of our candidates with the previously confirmed planets from \gls{ExoFOP} and expected planetary detections for \gls{TESS}. The following sections detail this property comparison. Generally, our candidates have properties consistent with the confirmed and expected planet distributions.

\begin{figure}
    \centering
    \includegraphics[width=\columnwidth]
    {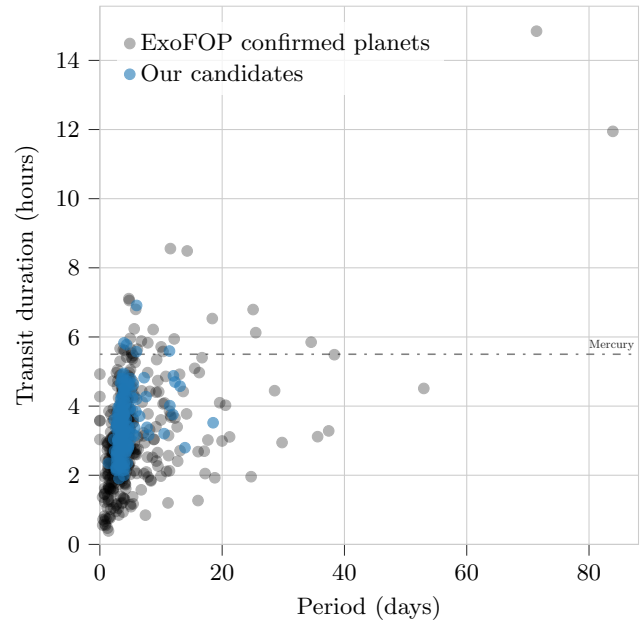}
    \caption{The distribution of the period and transit duration of the planet candidates.}
    \label{fig:candidate_transit_parameter_plot}
\end{figure}

\subsection{Conditional candidates}\label{sec:conditional-candidates}
Often, a potential transit signal will have an ambiguous source due to the proximity of two or more potential sources. Most signals where the source target is ambiguous are not included as candidates in our list (see \cref{subsec:post-processing}). The exception to this is when the signal would result in a planet candidate regardless of which of the ambiguous sources the signal originates from. Of our planet candidates, there are \numberOfConditionalCandidates{} for which there are two potential host star targets, where the signal coming from either source would suggest a radius consistent with a planet. These candidate source pairs are \conditionalPairsList{}. Our total count of candidates includes these \numberOfConditionalCandidates{} candidates. \cref{tab:planet_candidate_table} includes all \numberOfConditionalTargets{} ambiguous source targets, and the parameters assuming the candidate is from that source target. All other figures exclude these ambiguous targets.
     %\section{Discussion of planet candidates}\label{subsec:discussion-of-planet-candidates}
\subsection{Comparison with the ExoFOP confirmed planets}\label{subsubsec:comparison-with-exofop}
\begin{figure*}
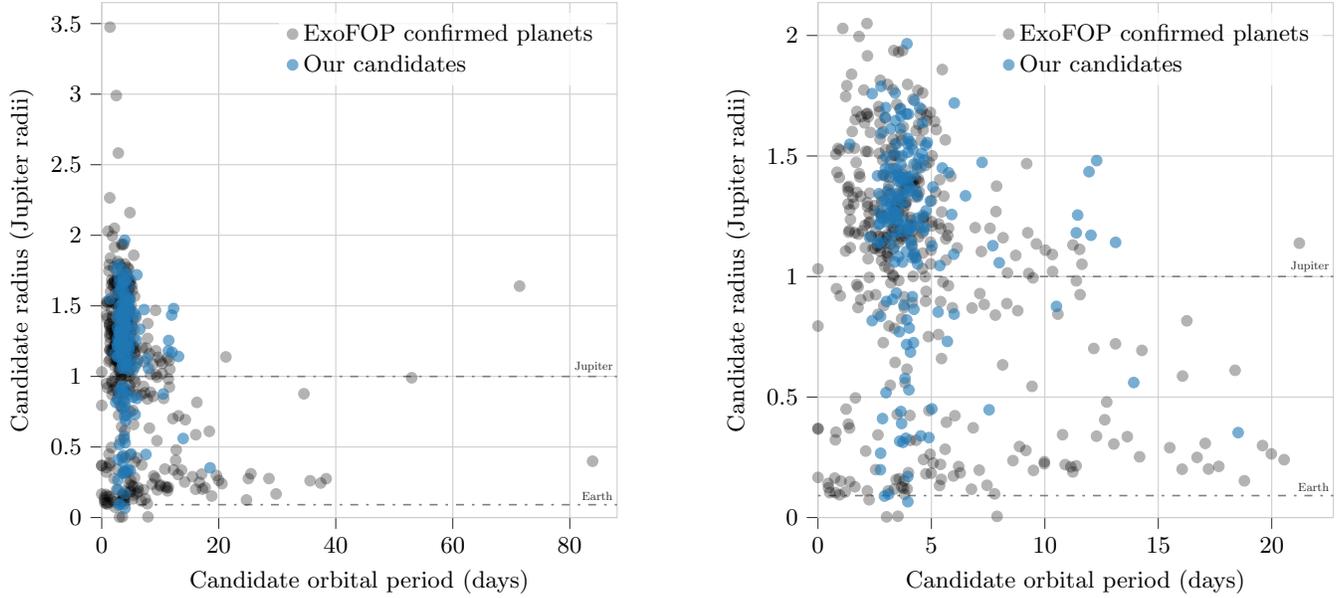

    \centering
    \subfloat{
        \includegraphics[width=0.46\textwidth]{candidate_radius_vs_orbital_period_plot.tikz}
    }
    \hfill
    \subfloat{
        \includegraphics[width=0.46\textwidth]{candidate_radius_vs_orbital_period_plot_zoomed.tikz}
    }
    \caption{Comparison between the distribution of the orbital periods and planetary radii of our planet candidates and ExoFOP confirmed planets (left panel); zoomed version (right panel). The candidates in this graph have their properties displayed in \cref{tab:planet_candidate_table}.}
    \label{fig:cadidate_radius_orbitalperiod_plot}
\end{figure*}
\Glsentryfull{ExoFOP} provides follow-up studies of targets observed by \gls{TESS}. \gls{ExoFOP} uses the stellar parameters from the \glsentryfull{TIC} and planet parameters from the NASA Exoplanet Archive~\citep{akeson2013nasa}. As our network is trained using the \glspl{LC} of targets with \gls{ExoFOP} confirmed planets, we expect our candidates' \glspl{LC} to exhibit similar features; the network learns that these features correspond to planetary transit signals. When these \gls{LC} features correspond to planet and/or star properties, we expect the characteristics of our candidates to be similar.

For example, the majority of \gls{ExoFOP} confirmed planets have a period of less than 5~days. As such, we expect our network to be inclined to search for planets with similar periods. Indeed, this tendency is observed in \cref{fig:cadidate_radius_orbitalperiod_plot}. This trend might likely be due not only to the training data distribution but also to the shorter period resulting in more transiting events in a single \gls{LC}, which likely makes the candidate easier to detect. The \glspl{LC} used by the NN are single \gls{TESS} sector \glspl{LC}, which have observing periods of \approximately27 days. This aspect of the data results in the majority of the transits identified having a relatively short period, as these are the only cases where multiple transits can be observed within a single sector. While there is nothing that explicitly restricts the network from labeling a \gls{LC} as a candidate even if it only contains a single transit event, non-periodic events are likely discouraged by the training process due to the possibility that they are caused more frequently by non-planetary sources. We expect the network to likely be more confident about signals with many periods being exhibited. This, combined with the majority of training samples being of short period, probably plays a factor in the network's decisions. At the same time, \cref{fig:cadidate_radius_orbitalperiod_plot} shows that the network does not seem to overemphasize predicting candidates with longer duration transits.

In addition to orbital period, \cref{fig:cadidate_radius_orbitalperiod_plot} also shows the radii of our candidates and \gls{ExoFOP} confirmed planets. Similar to the case of the periods, our candidates show a similar distribution of radii when compared to the \gls{ExoFOP} confirmed planets. The majority of candidates have a radius larger than Jupiter, with a smaller number of candidates having a radius between Jupiter and Earth. When observing the smallest radius candidates in \cref{fig:cadidate_radius_orbitalperiod_plot}, we again note the potential for spurious small radius estimates (see \cref{subsec:post-processing}).

\begin{figure}
    \centering
    \includegraphics[width=\columnwidth]
    {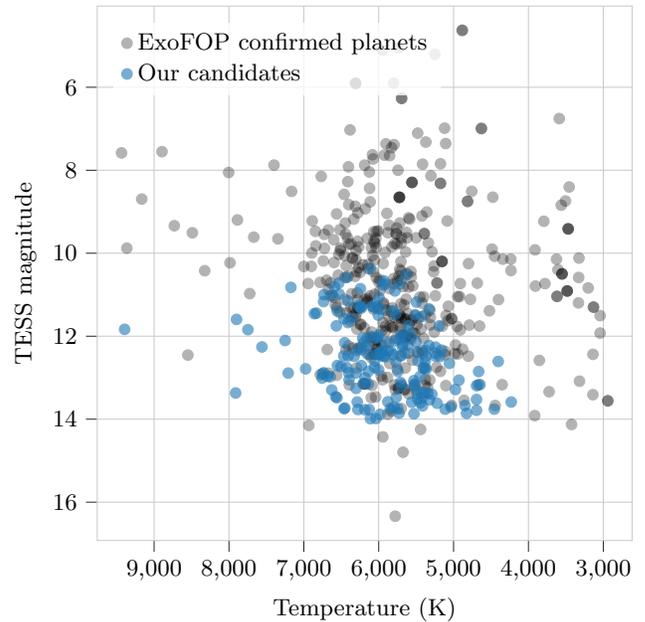}
    \caption{Color-magnitude diagram, showing the effective temperature of the host star vs.~its \gls{TESS} magnitude. Planet candidates from the present article are shown in blue, while those from~\gls{ExoFOP} are in gray.}
    \label{fig:color_magnitude_diagram}
\end{figure}
\cref{fig:color_magnitude_diagram} shows a color-magnitude diagram of the host stars of our candidates and the \gls{ExoFOP} confirmed planets. The range in \gls{TESS} magnitudes of the host stars of the planet candidates from our work is comparable with the ones of the host stars of planets confirmed by \gls{ExoFOP}. The host stars of our candidates have a higher magnitude (are less bright) than the host stars of the \gls{ExoFOP} confirmed planets. This is expected, as \gls{TESS} selects relatively bright targets for 2-minute cadence observing~\citep{ricker2014transiting} and the \gls{TESS} team's FFI search (the Quick Look Pipeline, \texttt{QLP}) only searched for planets down to a magnitude of 13.5~\citep{huang2020photometry}. In comparison, our network uses the \gls{TESS} FFI \glspl{LC}, which include dimmer stars. Indeed, the number of potential targets increases exponentially as the magnitude increases. However, we do not expect candidates to increase exponentially with the number of targets, as transit events become more difficult to detect around dimmer targets, whose signals are relatively more contaminated with sources of noise.

\begin{figure}
    \centering
    \includegraphics[width=\columnwidth]
    {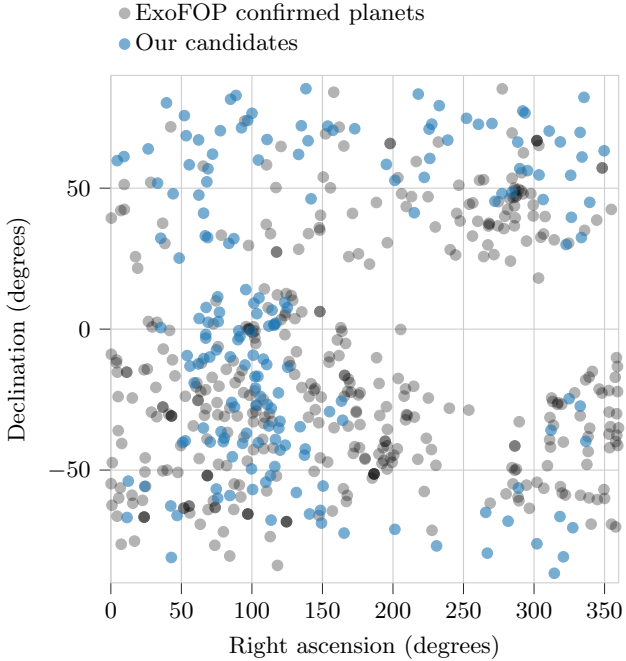}
    \caption{Position on the sky of our candidates compared with the \gls{ExoFOP} confirmed planets.}
    \label{fig:sky_coordinates}
\end{figure}

By comparing the position in the sky of our candidates for host stars with those confirmed by \gls{ExoFOP}, we note an homogeneous distribution of candidates across the sky (no region with strong preference). This comparison is shown in \cref{fig:sky_coordinates}. This is expected, as the network design does not take into account sky position and has no mechanisms designed to specifically prefer any region of the sky. One bias we expect from the network with regard to sky position would be a preference to candidate positions similar to those in confirmed \gls{ExoFOP} distribution, as these were the positions of the training examples. This may lead the network to prefer \glspl{LC} with sector specific noise from the sectors containing the most confirmed planets. However, to find candidates in similar regions to the confirmed planets may also simply be due these regions having clearer signals and less noise, in which case the reason for candidates being in similar regions may be the result of data quality rather than network bias. As the \gls{FFI} \glspl{LC} including dimmer magnitudes than those of the \gls{ExoFOP} confirmed planets, we might expect the \gls{NN} to prefer less crowded areas of the sky, where low brightness targets will have a high signal-to-noise ratio. Regardless, the network does not show any significant region omissions compared to the confirmed \gls{ExoFOP} planets.

\subsection{Comparison with the expected candidates}\label{subsubsec:comparison-with-expected}
\begin{figure}
    \centering
    \includegraphics[width=\columnwidth]
    {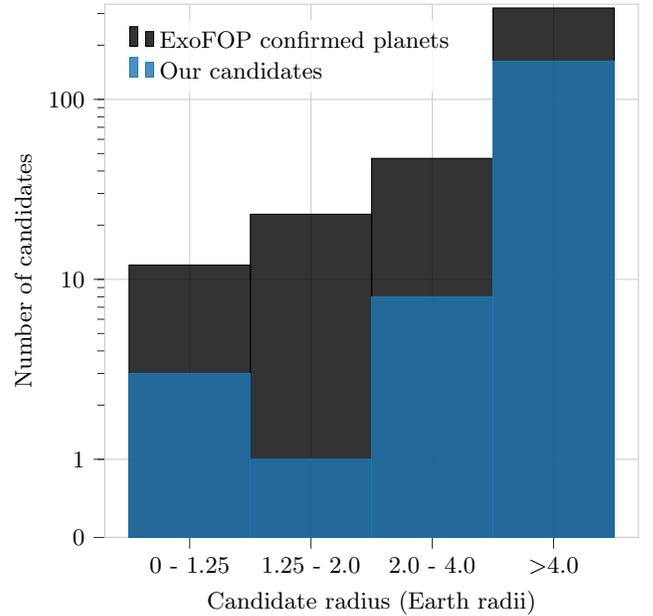}
    \caption{Planet radii distribution of our candidates compared with the confirmed \gls{ExoFOP} planets.
    The data are binned as follows: $<1.25$, $1.25-2.0$, $2.0-4.0$, $>4.0$ Earth radii.}
    \label{fig:candidate_earthradii_distribution_plot}
\end{figure}
\begin{figure}
    \centering
    \includegraphics[width=\columnwidth]
    {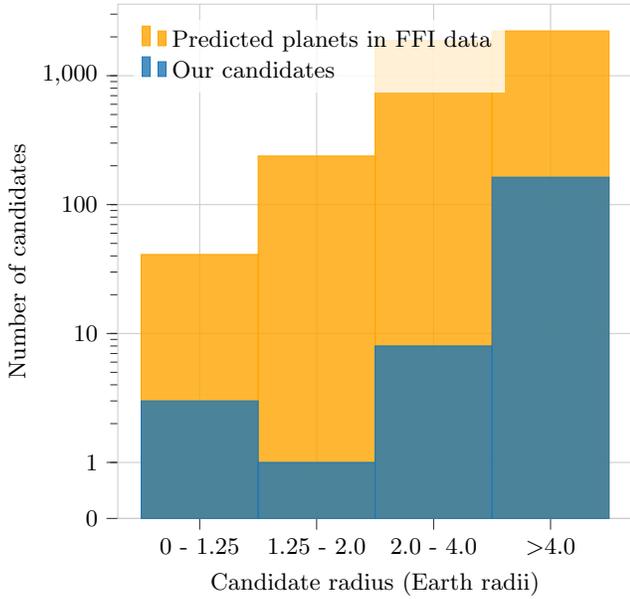}
    \caption{Planet radii distribution of our candidates compared with the expected findings predicted for the FFI data by \citet{barclay2018revised}.
    The data are binned as follows: $<1.25$, $1.25-2.0$, $2.0-4.0$, $>4.0$ Earth radii.}
    \label{fig:candidate_earthradii_distribution_comparison_B18_plot}
\end{figure}
\begin{figure}
    \centering
    \includegraphics[width=\columnwidth]
    {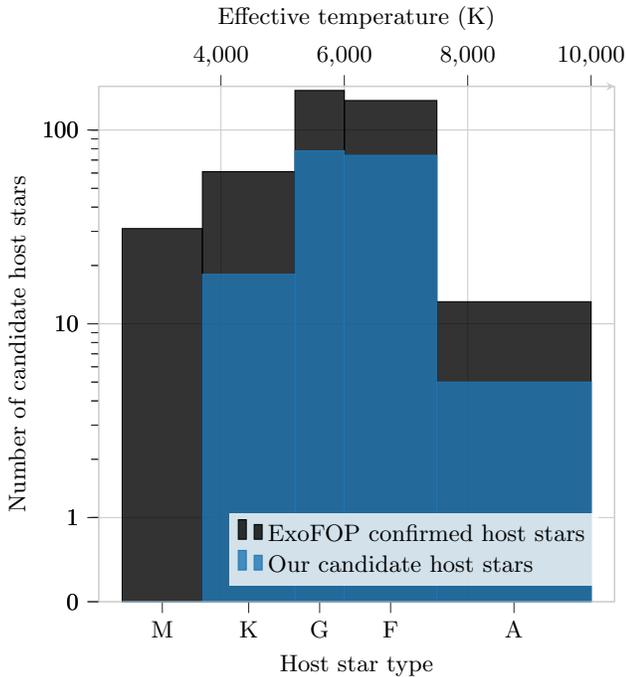}
    \caption{Stellar types of our candidates' hosts compared to the confirmed \gls{ExoFOP} planet hosts. The effective temperature of the stellar host is shown on the x-axis, shown also as spectral type.}
    \label{fig:candidate_stellar_temperature_plot}
\end{figure}

\begin{figure}
    \centering
    \includegraphics[width=\columnwidth]
    {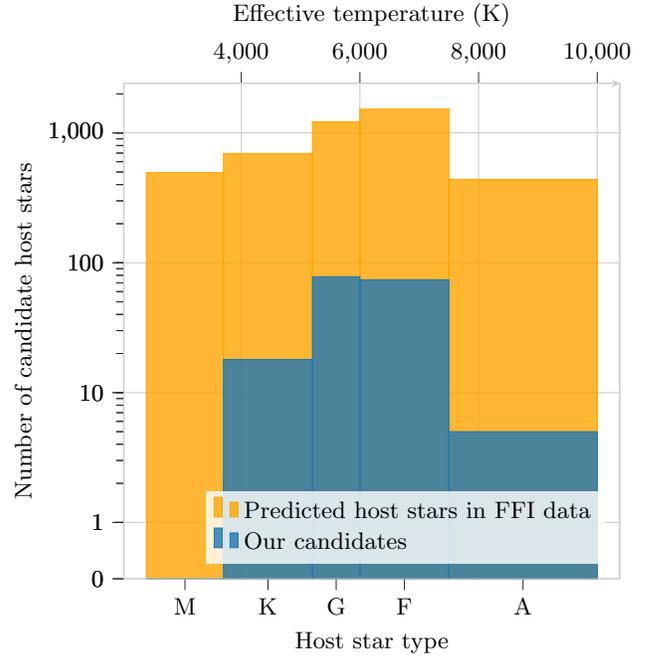}
    \caption{Stellar types of our candidates' hosts compared to the predictions for the FFI data by \citet{barclay2018revised}. The effective temperature of the stellar host is shown on the x-axis, shown also as spectral type.}
    \label{fig:candidate_stellar_temperature_comparison_B18_plot}
\end{figure}

The majority of \gls{ExoFOP} confirmed planets were found using \gls{TESS} 2-minute cadence data~\citep{huang2020photometry}. While \citet{barclay2018revised} expected the FFI \glspl{LC} to lead to proportionally higher radius planet discoveries compared to the 2-minute cadence \glspl{LC}, we do not see a significant difference in the distributions of our candidates compared to the \gls{ExoFOP} confirmed planets, as shown in \cref{fig:candidate_earthradii_distribution_plot}. This may be in part because the network is trained to find candidates similar to those in the training dataset. However, more likely is that larger planets are easier to find and confirm, and most of the existing \gls{ExoFOP} confirmed planets come from the higher end of the expectations of \citet{barclay2018revised}. A comparison of the distribution of our candidates to the expectations presented by \citet{barclay2018revised} is shown in \cref{fig:candidate_earthradii_distribution_comparison_B18_plot}.

\citet{barclay2018revised} expected that the majority of planets found in \gls{TESS} \gls{FFI} data would orbit G- and F-type stars. This is consistent with our finding, as shown in \cref{fig:candidate_stellar_temperature_plot}. This trend is true for the existing \gls{ExoFOP} confirmed planets as well. Notably, our results contain no M-type star hosts, and relatively few A-type star hosts. While planet candidates around these hosts are expected to be relatively rare in \gls{TESS} data and are relatively rare in the training data, our candidates are disproportionally low in these categories. This disparity is likely primarily caused by these categories having relatively few training examples. The \gls{ExoFOP} confirmed planets with M-type hosts have relatively smaller transit depths and shorter periods, which may explain the dearth of such candidates identified by our network.

\begin{figure}
    \centering
    \includegraphics[width=\columnwidth]
    {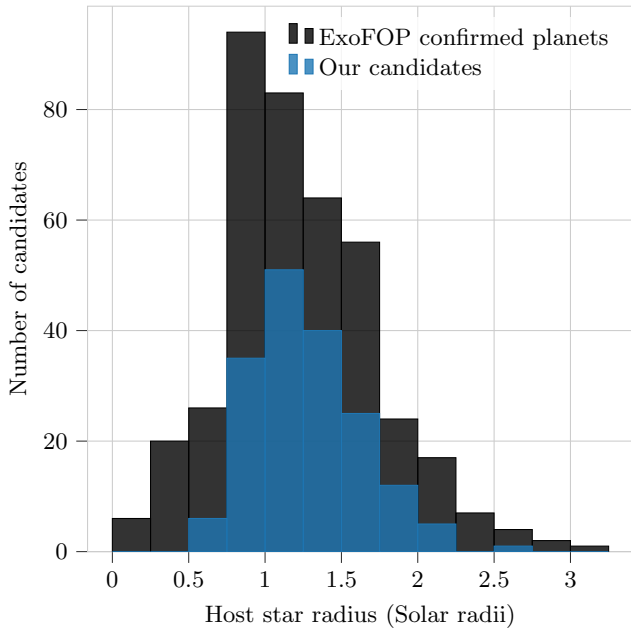}
    \caption{Stellar radii of our candidates' host stars compared to the ExoFOP confirmed host stars.}
    \label{fig:host_radii_distribution_plot}
\end{figure}
\citet{barclay2018revised} predicts that 80\% of planets found in \glspl{FFI} data are expected to orbit stars with radii larger than the Sun. \cref{fig:host_radii_distribution_plot} shows the distribution of our candidate host stars, where approximately 74\% are larger than the Sun.
% 118/160 = 73.75%

     \section{Conclusion}\label{sec:conclusion}
We present our convolutional \gls{NN}, which we train to identify planetary transit signals and dismiss false positives. To make a prediction for a given \gls{LC}, our network requires no prior transit parameters identified using other methods. We train our network using a dataset of confirmed exoplanets. Additionally, the network is trained to dismiss eclipsing binaries using a dataset of eclipsing binary candidates. We explain several network mechanisms and training techniques used to promote generalization of inference, including a method of injecting LCs into other LCs to create more varied training examples. Our network performs inference on a \gls{TESS} 30-minute cadence \gls{LC} in \approximately5ms on a single GPU, enabling large scale archival searches. We describe our post-identification analysis used to estimate transiter physical parameters. We present \numberOfHumanVettedCandidates{} new planet candidates identified by our network, which have passed subsequent human vetting designed to rule out false positives. We provide population analysis of our planet candidates and their host stars compared to a set of confirmed planets and the expected yield from \gls{TESS}. We provide to the public our \gls{NN} model as open-source code for further use and extension.
     \section{Acknowledgments}

This paper includes data collected by the {\em TESS} mission, which are publicly available from the Mikulski Archive for Space Telescopes (MAST). Funding for the {\em TESS} mission is provided by NASA's Science Mission directorate.

Resources supporting this work were provided by the NASA High-End Computing (HEC) Program through the NASA Center for Climate Simulation (NCCS) at Goddard Space Flight Center.

This research has made use of the Exoplanet Follow-up Observation Program website, which is operated by the California Institute of Technology, under contract with the National Aeronautics and Space Administration under the Exoplanet Exploration Program.

This work has made use of data from the European Space Agency (ESA) mission \textit{Gaia} (\url{https://www.cosmos.esa.int/gaia}), processed by the \textit{Gaia} Data Processing and Analysis Consortium (DPAC, \url{https://www.cosmos.esa.int/web/gaia/dpac/consortium}).

The material is based upon work supported by NASA under award number 80GSFC17M0002.

This research was supported by an appointment to the NASA Postdoctoral Program at the NASA Goddard Space Flight Center, administered by Universities Space Research Association under contract with NASA.

\facilities{
    \textit{Gaia},
    MAST,
    NCCS,
    \textit{TESS}
}

\software{
    \texttt{Astropy}~\citep{astropy2013,astropy2018},
    \texttt{Bokeh}~\citep{bokeh},
    \texttt{Eleanor}~\citep{feinstein2019eleanor},
    \texttt{Keras}~\citep{keras},
    \texttt{Lightkurve}~\citep{lightkurve},
    \texttt{Matplotlib}~\citep{matplotlib},
    \texttt{NumPy}~\citep{numpy},
    \texttt{Pandas}~\citep{pandas},
    \texttt{pytest}~\citep{pytestx.y},
    \texttt{Python}~\citep{python38},
    \texttt{Tensorflow}~\citep{tensorflow}
}
     \FloatBarrier
\bibliography{bibliography.bib}

\begin{thebibliography}{}
\expandafter\ifx\csname natexlab\endcsname\relax\def\natexlab#1{#1}\fi
\providecommand{\url}[1]{\href{#1}{#1}}
\providecommand{\dodoi}[1]{doi:~\href{http://doi.org/#1}{\nolinkurl{#1}}}
\providecommand{\doeprint}[1]{\href{http://ascl.net/#1}{\nolinkurl{http://ascl.net/#1}}}
\providecommand{\doarXiv}[1]{\href{https://arxiv.org/abs/#1}{\nolinkurl{https://arxiv.org/abs/#1}}}

\bibitem[{Abadi {et~al.}(2015)Abadi, Agarwal, Barham, Brevdo, Chen, Citro,
  Corrado, Davis, Dean, Devin, Ghemawat, Goodfellow, Harp, Irving, Isard, Jia,
  Jozefowicz, Kaiser, Kudlur, Levenberg, Man\'{e}, Monga, Moore, Murray, Olah,
  Schuster, Shlens, Steiner, Sutskever, Talwar, Tucker, Vanhoucke, Vasudevan,
  Vi\'{e}gas, Vinyals, Warden, Wattenberg, Wicke, Yu, \& Zheng}]{tensorflow}
Abadi, M., Agarwal, A., Barham, P., {et~al.} 2015, {TensorFlow}: Large-Scale
  Machine Learning on Heterogeneous Systems.
\newblock \url{https://www.tensorflow.org/}

\bibitem[{Akeson {et~al.}(2013)Akeson, Chen, Ciardi, Crane, Good, Harbut,
  Jackson, Kane, Laity, Leifer, {et~al.}}]{akeson2013nasa}
Akeson, R., Chen, X., Ciardi, D., {et~al.} 2013, Publications of the
  Astronomical Society of the Pacific, 125, 989

\bibitem[{Armstrong {et~al.}(2016)Armstrong, Pollacco, \&
  Santerne}]{armstrong2016transit}
Armstrong, D.~J., Pollacco, D., \& Santerne, A. 2016, Monthly Notices of the
  Royal Astronomical Society, 465, 2634, \dodoi{10.1093/mnras/stw2881}

\bibitem[{{Astropy Collaboration} {et~al.}(2013){Astropy Collaboration},
  {Robitaille}, {Tollerud}, {Greenfield}, {Droettboom}, {Bray}, {Aldcroft},
  {Davis}, {Ginsburg}, {Price-Whelan}, {Kerzendorf}, {Conley}, {Crighton},
  {Barbary}, {Muna}, {Ferguson}, {Grollier}, {Parikh}, {Nair}, {Unther},
  {Deil}, {Woillez}, {Conseil}, {Kramer}, {Turner}, {Singer}, {Fox}, {Weaver},
  {Zabalza}, {Edwards}, {Azalee Bostroem}, {Burke}, {Casey}, {Crawford},
  {Dencheva}, {Ely}, {Jenness}, {Labrie}, {Lim}, {Pierfederici}, {Pontzen},
  {Ptak}, {Refsdal}, {Servillat}, \& {Streicher}}]{astropy2013}
{Astropy Collaboration}, {Robitaille}, T.~P., {Tollerud}, E.~J., {et~al.} 2013,
  \aap, 558, A33, \dodoi{10.1051/0004-6361/201322068}

\bibitem[{{Astropy Collaboration} {et~al.}(2018){Astropy Collaboration},
  {Price-Whelan}, {Sip{\H{o}}cz}, {G{\"u}nther}, {Lim}, {Crawford}, {Conseil},
  {Shupe}, {Craig}, {Dencheva}, {Ginsburg}, {Vand erPlas}, {Bradley},
  {P{\'e}rez-Su{\'a}rez}, {de Val-Borro}, {Aldcroft}, {Cruz}, {Robitaille},
  {Tollerud}, {Ardelean}, {Babej}, {Bach}, {Bachetti}, {Bakanov}, {Bamford},
  {Barentsen}, {Barmby}, {Baumbach}, {Berry}, {Biscani}, {Boquien}, {Bostroem},
  {Bouma}, {Brammer}, {Bray}, {Breytenbach}, {Buddelmeijer}, {Burke},
  {Calderone}, {Cano Rodr{\'\i}guez}, {Cara}, {Cardoso}, {Cheedella}, {Copin},
  {Corrales}, {Crichton}, {D'Avella}, {Deil}, {Depagne}, {Dietrich}, {Donath},
  {Droettboom}, {Earl}, {Erben}, {Fabbro}, {Ferreira}, {Finethy}, {Fox},
  {Garrison}, {Gibbons}, {Goldstein}, {Gommers}, {Greco}, {Greenfield},
  {Groener}, {Grollier}, {Hagen}, {Hirst}, {Homeier}, {Horton}, {Hosseinzadeh},
  {Hu}, {Hunkeler}, {Ivezi{\'c}}, {Jain}, {Jenness}, {Kanarek}, {Kendrew},
  {Kern}, {Kerzendorf}, {Khvalko}, {King}, {Kirkby}, {Kulkarni}, {Kumar},
  {Lee}, {Lenz}, {Littlefair}, {Ma}, {Macleod}, {Mastropietro}, {McCully},
  {Montagnac}, {Morris}, {Mueller}, {Mumford}, {Muna}, {Murphy}, {Nelson},
  {Nguyen}, {Ninan}, {N{\"o}the}, {Ogaz}, {Oh}, {Parejko}, {Parley}, {Pascual},
  {Patil}, {Patil}, {Plunkett}, {Prochaska}, {Rastogi}, {Reddy Janga},
  {Sabater}, {Sakurikar}, {Seifert}, {Sherbert}, {Sherwood-Taylor}, {Shih},
  {Sick}, {Silbiger}, {Singanamalla}, {Singer}, {Sladen}, {Sooley},
  {Sornarajah}, {Streicher}, {Teuben}, {Thomas}, {Tremblay}, {Turner},
  {Terr{\'o}n}, {van Kerkwijk}, {de la Vega}, {Watkins}, {Weaver}, {Whitmore},
  {Woillez}, {Zabalza}, \& {Astropy Contributors}}]{astropy2018}
{Astropy Collaboration}, {Price-Whelan}, A.~M., {Sip{\H{o}}cz}, B.~M., {et~al.}
  2018, \aj, 156, 123, \dodoi{10.3847/1538-3881/aabc4f}

\bibitem[{Barclay {et~al.}(2018)Barclay, Pepper, \&
  Quintana}]{barclay2018revised}
Barclay, T., Pepper, J., \& Quintana, E.~V. 2018, The Astrophysical Journal
  Supplement Series, 239, 2

\bibitem[{{Bokeh Development Team}(2020)}]{bokeh}
{Bokeh Development Team}. 2020, Bokeh: Python library for interactive
  visualization.
\newblock \url{https://bokeh.org/}

\bibitem[{Brown {et~al.}(2018)Brown, Vallenari, Prusti, De~Bruijne, Babusiaux,
  Bailer-Jones, Biermann, Evans, Eyer, Jansen, {et~al.}}]{brown2018gaia}
Brown, A., Vallenari, A., Prusti, T., {et~al.} 2018, Astronomy \& astrophysics,
  616, A1

\bibitem[{{Burke} {et~al.}(2020){Burke}, {Levine}, {Fausnaugh}, {Vanderspek},
  {Barclay}, {Libby-Roberts}, {Morris}, {Sipocz}, {Owens}, {Feinstein}, \&
  {Camacho}}]{burke2020tess}
{Burke}, C.~J., {Levine}, A., {Fausnaugh}, M., {et~al.} 2020, {TESS-Point: High
  precision TESS pointing tool}.
\newblock \doeprint{2003.001}

\bibitem[{Carmichael {et~al.}(2020)Carmichael, Quinn, Mustill, Huang, Zhou,
  Persson, Nielsen, Collins, Ziegler, Collins, {et~al.}}]{carmichael2020two}
Carmichael, T.~W., Quinn, S.~N., Mustill, A.~J., {et~al.} 2020, The
  Astronomical Journal, 160, 53

\bibitem[{Chollet {et~al.}(2015)}]{keras}
Chollet, F., {et~al.} 2015, Keras, \url{https://keras.io}

\bibitem[{Crouzet {et~al.}(2017)Crouzet, McCullough, Long, Rodriguez, des
  Etangs, Ribas, Bourrier, H{\'e}brard, Vilardell, Deleuil,
  {et~al.}}]{crouzet2017discovery}
Crouzet, N., McCullough, P., Long, D., {et~al.} 2017, The Astronomical Journal,
  153, 94

\bibitem[{Cybenko(1989)}]{cybenko1989approximation}
Cybenko, G. 1989, Mathematics of control, signals and systems, 2, 303

\bibitem[{Dong {et~al.}(2014)Dong, Loy, He, \& Tang}]{dong2014learning}
Dong, C., Loy, C.~C., He, K., \& Tang, X. 2014, in European Conference on
  Computer Vision, Springer, 184--199

\bibitem[{Eastman {et~al.}(2010)Eastman, Siverd, \&
  Gaudi}]{eastman2010achieving}
Eastman, J., Siverd, R., \& Gaudi, B.~S. 2010, Publications of the Astronomical
  Society of the Pacific, 122, 935

\bibitem[{Feinstein {et~al.}(2019)Feinstein, Montet, Foreman-Mackey, Bedell,
  Saunders, Bean, Christiansen, Hedges, Luger, Scolnic,
  {et~al.}}]{feinstein2019eleanor}
Feinstein, A.~D., Montet, B.~T., Foreman-Mackey, D., {et~al.} 2019,
  Publications of the Astronomical Society of the Pacific, 131, 094502

\bibitem[{Foreman-Mackey {et~al.}(2017)Foreman-Mackey, Agol, Ambikasaran, \&
  Angus}]{foreman2017fast}
Foreman-Mackey, D., Agol, E., Ambikasaran, S., \& Angus, R. 2017, The
  Astronomical Journal, 154, 220

\bibitem[{Glorot \& Bengio(2010)}]{glorot2010understanding}
Glorot, X., \& Bengio, Y. 2010, in Proceedings of Machine Learning Research,
  Vol.~9, Proceedings of the Thirteenth International Conference on Artificial
  Intelligence and Statistics, ed. Y.~W. Teh \& M.~Titterington (Chia Laguna
  Resort, Sardinia, Italy: PMLR), 249--256.
\newblock \url{http://proceedings.mlr.press/v9/glorot10a.html}

\bibitem[{Glorot {et~al.}(2011)Glorot, Bordes, \& Bengio}]{glorot2011deep}
Glorot, X., Bordes, A., \& Bengio, Y. 2011, in Proceedings of Machine Learning
  Research, Vol.~15, Proceedings of the Fourteenth International Conference on
  Artificial Intelligence and Statistics, ed. G.~Gordon, D.~Dunson, \&
  M.~Dudík (Fort Lauderdale, FL, USA: PMLR), 315--323.
\newblock \url{http://proceedings.mlr.press/v15/glorot11a.html}

\bibitem[{Harris {et~al.}(2020)Harris, Millman, van~der Walt, Gommers,
  Virtanen, Cournapeau, Wieser, Taylor, Berg, Smith, Kern, Picus, Hoyer, van
  Kerkwijk, Brett, Haldane, del R{\'{i}}o, Wiebe, Peterson,
  G{\'{e}}rard-Marchant, Sheppard, Reddy, Weckesser, Abbasi, Gohlke, \&
  Oliphant}]{numpy}
Harris, C.~R., Millman, K.~J., van~der Walt, S.~J., {et~al.} 2020, Nature, 585,
  357, \dodoi{10.1038/s41586-020-2649-2}

\bibitem[{Hinton {et~al.}(2012)Hinton, Deng, Yu, Dahl, Mohamed, Jaitly, Senior,
  Vanhoucke, Nguyen, Sainath, {et~al.}}]{hinton2012deep}
Hinton, G., Deng, L., Yu, D., {et~al.} 2012, IEEE Signal Processing Magazine,
  29, 82

\bibitem[{Hochreiter(1998)}]{hochreiter1998vanishing}
Hochreiter, S. 1998, International Journal of Uncertainty, Fuzziness and
  Knowledge-Based Systems, 6, 107

\bibitem[{Huang {et~al.}(2020)Huang, Vanderburg, P{\'{a}}l, Sha, Yu, Fong,
  Fausnaugh, Shporer, Guerrero, Vanderspek, \& Ricker}]{huang2020photometry}
Huang, C.~X., Vanderburg, A., P{\'{a}}l, A., {et~al.} 2020, Research Notes of
  the {AAS}, 4, 206, \dodoi{10.3847/2515-5172/abca2d}

\bibitem[{Hunter(2007)}]{matplotlib}
Hunter, J.~D. 2007, Computing in science \& engineering, 9, 90

\bibitem[{Ioffe \& Szegedy(2015)}]{ioffe2015batch}
Ioffe, S., \& Szegedy, C. 2015, in Proceedings of Machine Learning Research,
  Vol.~37, Proceedings of the 32nd International Conference on Machine
  Learning, ed. F.~Bach \& D.~Blei (Lille, France: PMLR), 448--456.
\newblock \url{http://proceedings.mlr.press/v37/ioffe15.html}

\bibitem[{Jenkins {et~al.}(2016)Jenkins, Twicken, McCauliff, Campbell,
  Sanderfer, Lung, Mansouri-Samani, Girouard, Tenenbaum, Klaus,
  {et~al.}}]{jenkins2016tess}
Jenkins, J.~M., Twicken, J.~D., McCauliff, S., {et~al.} 2016, in Software and
  Cyberinfrastructure for Astronomy IV, Vol. 9913, International Society for
  Optics and Photonics, 99133E

\bibitem[{Kingma \& Ba(2015)}]{kingma2014adam}
Kingma, D.~P., \& Ba, J. 2015, in 3rd International Conference on Learning
  Representations, {ICLR} 2015, San Diego, CA, USA, May 7-9, 2015, Conference
  Track Proceedings, ed. Y.~Bengio \& Y.~LeCun.
\newblock \url{http://arxiv.org/abs/1412.6980}

\bibitem[{Kostov {et~al.}(2019)Kostov, Mullally, Quintana, Coughlin, Mullally,
  Barclay, Col{\'o}n, Schlieder, Barentsen, \& Burke}]{kostov2019discovery}
Kostov, V.~B., Mullally, S.~E., Quintana, E.~V., {et~al.} 2019, The
  Astronomical Journal, 157, 124

\bibitem[{Kov{\'a}cs {et~al.}(2002)Kov{\'a}cs, Zucker, \&
  Mazeh}]{kovacs2002box}
Kov{\'a}cs, G., Zucker, S., \& Mazeh, T. 2002, Astronomy and Astrophysics, 391,
  369

\bibitem[{Krekel {et~al.}(2004)Krekel, Oliveira, Pfannschmidt, Bruynooghe,
  Laugher, \& Bruhin}]{pytestx.y}
Krekel, H., Oliveira, B., Pfannschmidt, R., {et~al.} 2004, pytest x.y.
\newblock \url{https://github.com/pytest-dev/pytest}

\bibitem[{Krizhevsky {et~al.}(2012)Krizhevsky, Sutskever, \&
  Hinton}]{krizhevsky2012imagenet}
Krizhevsky, A., Sutskever, I., \& Hinton, G.~E. 2012, in Advances in Neural
  Information Processing Systems, ed. F.~Pereira, C.~J.~C. Burges, L.~Bottou,
  \& K.~Q. Weinberger, Vol.~25 (Curran Associates, Inc.), 1097--1105.
\newblock
  \url{https://proceedings.neurips.cc/paper/2012/file/c399862d3b9d6b76c8436e924a68c45b-Paper.pdf}

\bibitem[{Kruse {et~al.}(2019)Kruse, Agol, Luger, \&
  Foreman-Mackey}]{kruse2019detection}
Kruse, E., Agol, E., Luger, R., \& Foreman-Mackey, D. 2019, The Astrophysical
  Journal Supplement Series, 244, 11

\bibitem[{LeCun {et~al.}(2015)LeCun, Bengio, \& Hinton}]{lecun2015deep}
LeCun, Y., Bengio, Y., \& Hinton, G. 2015, nature, 521, 436

\bibitem[{LeCun {et~al.}(2012)LeCun, Bottou, Orr, \&
  M{\"u}ller}]{lecun2012efficient}
LeCun, Y.~A., Bottou, L., Orr, G.~B., \& M{\"u}ller, K.-R. 2012, in Neural
  networks: Tricks of the trade (Springer), 9--48

\bibitem[{Leshno {et~al.}(1993)Leshno, Lin, Pinkus, \&
  Schocken}]{leshno1993multilayer}
Leshno, M., Lin, V.~Y., Pinkus, A., \& Schocken, S. 1993, Neural networks, 6,
  861

\bibitem[{{Lightkurve Collaboration} {et~al.}(2018){Lightkurve Collaboration},
  {Cardoso}, {Hedges}, {Gully-Santiago}, {Saunders}, {Cody}, {Barclay}, {Hall},
  {Sagear}, {Turtelboom}, {Zhang}, {Tzanidakis}, {Mighell}, {Coughlin}, {Bell},
  {Berta-Thompson}, {Williams}, {Dotson}, \& {Barentsen}}]{lightkurve}
{Lightkurve Collaboration}, {Cardoso}, J.~V.~d.~M., {Hedges}, C., {et~al.}
  2018, {Lightkurve: Kepler and TESS time series analysis in Python}, 1.11.3,
  Astrophysics Source Code Library.
\newblock \doeprint{1812.013}

\bibitem[{Luger {et~al.}(2016)Luger, Agol, Kruse, Barnes, Becker,
  Foreman-Mackey, \& Deming}]{luger2016everest}
Luger, R., Agol, E., Kruse, E., {et~al.} 2016, The Astronomical Journal, 152,
  100

\bibitem[{{Mikulski Archive for Space Telescopes}(2020)}]{masttess}
{Mikulski Archive for Space Telescopes}. 2020, Mikulski Archive for Space
  Telescopes, \url{https://archive.stsci.edu/missions-and-data/tess}

\bibitem[{Olmschenk {et~al.}(2021)Olmschenk, Silva, Barry, \&
  Wyrwas}]{greg_olmschenk_2021_4460662}
Olmschenk, G., Silva, S.~I., Barry, R., \& Wyrwas, E. 2021, {golmschenk/ramjet:
  Identifying Planetary Transit Candidates in TESS Full-Frame Image Light
  Curves via Convolutional Neural Networks},
  2021-tess-ffi-planet-candidates-paper,  Zenodo,
  \dodoi{10.5281/zenodo.4460662}

\bibitem[{{Python Core Team}(2020)}]{python38}
{Python Core Team}. 2020, {Python: A dynamic, open source programming
  language}, {Python Software Foundation}.
\newblock \url{https://www.python.org/}

\bibitem[{Ricker {et~al.}(2014)Ricker, Winn, Vanderspek, Latham, Bakos, Bean,
  Berta-Thompson, Brown, Buchhave, Butler, {et~al.}}]{ricker2014transiting}
Ricker, G.~R., Winn, J.~N., Vanderspek, R., {et~al.} 2014, Journal of
  Astronomical Telescopes, Instruments, and Systems, 1, 014003

\bibitem[{Rumelhart {et~al.}(1986{\natexlab{a}})Rumelhart, Hinton, \&
  Williams}]{rumelhart1985learning}
Rumelhart, D.~E., Hinton, G.~E., \& Williams, R.~J. 1986{\natexlab{a}},
  Learning Internal Representations by Error Propagation (Cambridge, MA, USA:
  MIT Press), 318–362

\bibitem[{Rumelhart {et~al.}(1986{\natexlab{b}})Rumelhart, Hinton, \&
  Williams}]{rumelhart1986learning}
---. 1986{\natexlab{b}}, nature, 323, 533

\bibitem[{Srivastava {et~al.}(2014)Srivastava, Hinton, Krizhevsky, Sutskever,
  \& Salakhutdinov}]{srivastava2014dropout}
Srivastava, N., Hinton, G., Krizhevsky, A., Sutskever, I., \& Salakhutdinov, R.
  2014, The journal of machine learning research, 15, 1929

\bibitem[{Stassun {et~al.}(2018)Stassun, Oelkers, Pepper, Paegert, De~Lee,
  Torres, Latham, Charpinet, Dressing, Huber, {et~al.}}]{stassun2018tess}
Stassun, K.~G., Oelkers, R.~J., Pepper, J., {et~al.} 2018, The Astronomical
  Journal, 156, 102

\bibitem[{Sullivan {et~al.}(2015)Sullivan, Winn, Berta-Thompson, Charbonneau,
  Deming, Dressing, Latham, Levine, McCullough, Morton,
  {et~al.}}]{sullivan2015transiting}
Sullivan, P.~W., Winn, J.~N., Berta-Thompson, Z.~K., {et~al.} 2015, The
  Astrophysical Journal, 809, 77

\bibitem[{Tenenbaum \& Jenkins(2018)}]{tenenbaum2018tess}
Tenenbaum, P., \& Jenkins, J.~M. 2018, TESS Science Data Products Description
  Document, Tech. rep., EXP-TESS-ARC-ICD-0014 Rev D https://archive. stsci.
  edu/missions/tess/doc~…

\bibitem[{{TESS Follow-up Observing Program Working Group}(2020)}]{exofoptess}
{TESS Follow-up Observing Program Working Group}. 2020, The Exoplanet Follow-up
  Observing Program for TESS, \url{https://exofop.ipac.caltech.edu/tess/}

\bibitem[{Tompson {et~al.}(2015)Tompson, Goroshin, Jain, Lecun, \&
  Bregler}]{tompson2015efficient}
Tompson, J., Goroshin, R., Jain, A., Lecun, Y., \& Bregler, C. 2015, in
  Proceedings of the IEEE conference on Computer Vision and Pattern
  Recognition, 648--656, \dodoi{10.1109/CVPR.2015.7298664}

\bibitem[{{W}es {M}c{K}inney(2010)}]{pandas}
{W}es {M}c{K}inney. 2010, in {P}roceedings of the 9th {P}ython in {S}cience
  {C}onference, ed. {S}t\'efan van~der {W}alt \& {J}arrod {M}illman, 56 -- 61,
  \dodoi{10.25080/Majora-92bf1922-00a}

\bibitem[{Wilson \& Cowan(1972)}]{wilson1972excitatory}
Wilson, H.~R., \& Cowan, J.~D. 1972, Biophysical journal, 12, 1

\bibitem[{Xu {et~al.}(2014)Xu, Ren, Liu, \& Jia}]{xu2014deep}
Xu, L., Ren, J.~S., Liu, C., \& Jia, J. 2014, in Advances in Neural Information
  Processing Systems, ed. Z.~Ghahramani, M.~Welling, C.~Cortes, N.~Lawrence, \&
  K.~Q. Weinberger, Vol.~27 (Curran Associates, Inc.), 1790--1798.
\newblock
  \url{https://proceedings.neurips.cc/paper/2014/file/1c1d4df596d01da60385f0bb17a4a9e0-Paper.pdf}

\bibitem[{Zhong {et~al.}(2020)Zhong, Zheng, Kang, Li, \&
  Yang}]{zhong2020random}
Zhong, Z., Zheng, L., Kang, G., Li, S., \& Yang, Y. 2020, in Proceedings of the
  AAAI Conference on Artificial Intelligence, Vol.~34, 13001--13008,
  \dodoi{10.1609/aaai.v34i07.7000}

\bibitem[{Zhou(2020)}]{zhou2020universality}
Zhou, D.-X. 2020, Applied and computational harmonic analysis, 48, 787

\bibitem[{Zhou {et~al.}(2017)Zhou, Bakos, Hartman, Latham, Torres, Bhatti,
  Penev, Buchhave, Kov{\'a}cs, Bieryla, {et~al.}}]{zhou2017hat}
Zhou, G., Bakos, G., Hartman, J.~D., {et~al.} 2017, The Astronomical Journal,
  153, 211

\end{thebibliography}
\bibliographystyle{aasjournal}
\clearpage
     {
\centering
% WARNING! THIS FILE IS AUTO-GENERATED!
% Do not manually edit this file.
% Changes to this file will be overwritten by the next auto-generation.
% Edit the generating file instead.
% This file was generated by `results_resources/candidate_table_generator.py`.
\begin{longtable}{rrrrrr}
\caption{The list of human-vetted planet candidates by TIC ID with transit parameters. Candidate radius is
                   estimated based on several assumptions (see text for details). A `$^{\dagger}$'
                   mark signifies cases where the contamination ratio of the target is unknown and, in the radius
                   estimation, zero is used. A `$^{\ddagger}$' mark signifies cases where a radius could not be
                   estimated due to a lack of stellar parameters. A `$^{*}$' mark signifies that there is
                   ambiguity about whether this target or another marked with `$^{*}$' is the candidate
                   host (see \cref{sec:conditional-candidates}).}
\label{tab:planet_candidate_table}\\
\toprule
   TIC ID & $\begin{tabular}{@{}c@{}}Candidate radius\\(Jupiter radii)\end{tabular}$ & $\begin{tabular}{@{}c@{}}Transit epoch\\(BTJD)\end{tabular}$ & $\begin{tabular}{@{}c@{}}Orbital period\\(days)\end{tabular}$ & $\begin{tabular}{@{}c@{}}Transit duration\\(hours)\end{tabular}$ & $\begin{tabular}{@{}c@{}}Transit depth\\(ppm)\end{tabular}$ \\
\midrule
\endfirsthead
\caption[]{The list of human-vetted planet candidates by TIC ID with transit parameters. Candidate radius is
                   estimated based on several assumptions (see text for details). A `$^{\dagger}$'
                   mark signifies cases where the contamination ratio of the target is unknown and, in the radius
                   estimation, zero is used. A `$^{\ddagger}$' mark signifies cases where a radius could not be
                   estimated due to a lack of stellar parameters. A `$^{*}$' mark signifies that there is
                   ambiguity about whether this target or another marked with `$^{*}$' is the candidate
                   host (see \cref{sec:conditional-candidates}).} \\
\toprule
   TIC ID & $\begin{tabular}{@{}c@{}}Candidate radius\\(Jupiter radii)\end{tabular}$ & $\begin{tabular}{@{}c@{}}Transit epoch\\(BTJD)\end{tabular}$ & $\begin{tabular}{@{}c@{}}Orbital period\\(days)\end{tabular}$ & $\begin{tabular}{@{}c@{}}Transit duration\\(hours)\end{tabular}$ & $\begin{tabular}{@{}c@{}}Transit depth\\(ppm)\end{tabular}$ \\
\midrule
\endhead
\midrule
\multicolumn{6}{r}{{Continued on next page}} \\
\midrule
\endfoot

\bottomrule
\endlastfoot
  7548817 &                                        1.45\rlap{} &                                           1687.086 &                                             4.5207 &                                              4.597 &                                               8864 \\
  9443323 &                                        1.41\rlap{} &                                           1818.027 &                                             4.0135 &                                              3.923 &                                               7659 \\
 11755687 &                                        1.24\rlap{} &                                           1468.686 &                                             3.0525 &                                              2.308 &                                              15591 \\
 12090836 &                                        1.61\rlap{} &                                           1470.350 &                                             3.1972 &                                              3.740 &                                              10744 \\
 14173089 &                            0.69\rlap{$^{\dagger}$} &                                           1441.471 &                                             4.0756 &                                              3.113 &                                               3709 \\
 21279791 &                                        0.90\rlap{} &                                           1685.697 &                                             3.0183 &                                              3.232 &                                               7249 \\
 27414976 &                            1.39\rlap{$^{\dagger}$} &                                           1741.338 &                                             3.9765 &                                              2.339 &                                              32848 \\
\end{longtable}
 }
 \end{document}